\documentclass[journal=jpclcd,manuscript=article,articletitle=true,layout=twocolumn]{achemso}

\usepackage{amsmath}
\usepackage{amssymb}
\usepackage{latexsym}
\usepackage{graphicx}
\usepackage{epsfig,epsf,rotating}
\usepackage{indentfirst}
\usepackage{epstopdf}
\usepackage{xcolor}
\usepackage{multirow}
\usepackage{tabularx}
\usepackage{ctable}

\usepackage{upgreek}
\newcolumntype{C}{>{\centering\arraybackslash}X}
\usepackage{nicefrac}

\title{Effect of Nano-Confinement on NMR Relaxation of Heptane in Kerogen from MD Simulations and Measurements\footnote{Notice:  This manuscript has been authored by UT-Battelle, LLC, under contract DE-AC05-00OR22725 with the US Department of Energy (DOE). The US government retains and the publisher, by accepting the article for publication, acknowledges that the US government retains a nonexclusive, paid-up, irrevocable, worldwide license to publish or reproduce the published form of this manuscript, or allow others to do so, for US government purposes. DOE will provide public access to these results of federally sponsored research in accordance with the DOE Public Access Plan (http://energy.gov/downloads/doe-public-access-plan).}}
\author{Arjun Valiya Parambathu}
\affiliation{Department of Chemical and Biomolecular Engineering, Rice University, 6100 Main St., Houston, TX 77005, USA}
\alsoaffiliation{Department of Chemical and Biomolecular Engineering, 150 Academy St, University of Delaware, Newark, DE 19716, USA}
\author{Walter G. Chapman}
\affiliation{Department of Chemical and Biomolecular Engineering, Rice University, 6100 Main St., Houston, TX 77005, USA}
\author{George J. Hirasaki}
\affiliation{Department of Chemical and Biomolecular Engineering, Rice University, 6100 Main St., Houston, TX 77005, USA}
\author{Dilip~Asthagiri}
\affiliation{Oak Ridge National Laboratory, 1 Bethel Valley Road, Oak Ridge, TN 37830-6012}
\email{asthagiridn@ornl.gov}
\author{Philip M. Singer}
\affiliation{Department of Chemical and Biomolecular Engineering, Rice University, 6100 Main St., Houston, TX 77005, USA}
\email{ps41@rice.edu}

\begin{document}
	\begin{abstract}
	 Kerogen-rich shale reservoirs will play a key role during the energy transition, yet the effects of nano-confinement on the NMR relaxation of hydrocarbons in kerogen are poorly understood. We use atomistic MD simulations to investigate the effects of nano-confinement on the $^1$H NMR relaxation times $T_1$ and $T_2$ of heptane in kerogen. In the case of $T_1$, we discover the important role of confinement in reducing $T_1$ by $\sim$3 orders of magnitude from bulk heptane, in agreement with measurements of heptane dissolved in kerogen from the Kimmeridge Shale, \textit{without any models or free parameters}. In the case of $T_2$, we discover that confinement breaks spatial isotropy and gives rise to residual dipolar coupling which reduces $T_2$ by $\sim$5 orders of magnitude from bulk heptane. We use the simulated $T_2$ to calibrate the surface relaxivity and thence predict the pore-size distribution of the organic nano-pores in kerogen, \textit{without additional experimental data.} 
		
		\begin{tocentry} 
			\includegraphics{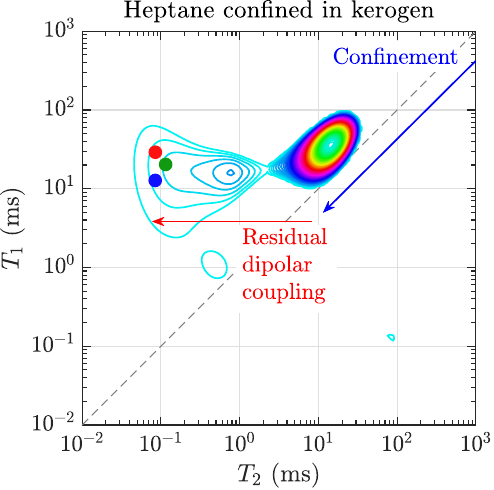} 
		\end{tocentry}
		
	\end{abstract}
	
	\maketitle
	
	
	Kerogen is a high molecular-weight organic matter that serves as source and reservoir for all hydrocarbons in ``shale''. Kerogen-rich shale reservoirs will most certainly play a key role during the energy transition, with it providing a clean burning energy source, as well as a sink for carbon capture and storage. This strategy requires us to have an accurate mapping of the subterranean shale formation, since the ``sweet spots" for economical extraction via fracking \cite{yethiraj:jpcl2013} and CO$_2$ sequestration \cite{hosseininoosheri2018} are an intimate function of the local structure of the extremely heterogeneous shale formations. Among the many tools used in characterizing the reservoirs, NMR (nuclear magnetic resonance) logging is at the forefront. 
	
	The interpretation of NMR logs still relies on traditional theories \cite{bloembergen:pr1948,torrey:pr1953} that make rather strong assumptions regarding molecular structure and interactions. For example, the seminal Bloemebergen, Purcell, Pound (BPP) theory for \textit{intra}molecular $^1$H-$^1$H dipole-dipole relaxation assumes molecules are hard spheres that undergo free rotation. This model is the basis for modeling NMR relaxation in liquids and predicts that $T_1$ increases with viscosity ($T_1 \propto \eta/T$) at high viscosity. However, for highly viscous or nanoconfined liquids, i.e.\ in the slow-motion region, experimental results  \cite{vinegar:spefe1991,latorraca:spwla1999,zhang:spwla2002,yang:jmr2008,yang:petro2012,kausik:petro2019,singer:EF2018,singer:jpcb2020} show that $T_1$ becomes independent of $\eta/T$. The literature commonly attributes this deviation by invoking the physics of paramagnetism \cite{serve2021}, or it is attributed to the presence of organic nano-pores in the system that confines the small molecules and enhances their $^1$H-$^1$H dipole-dipole relaxation \cite{singer:jpcb2020}.
	
	\begin{figure*}[ht!]
		\begin{center}
			\includegraphics{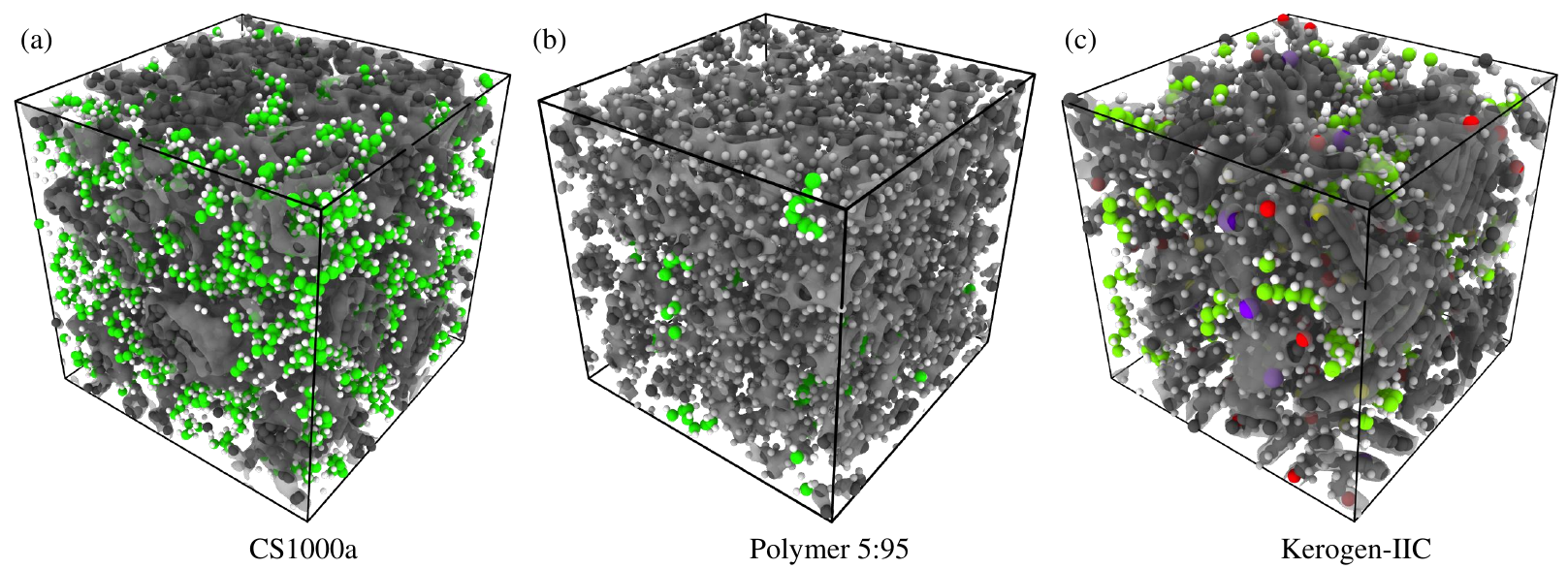}
		    \caption{Equilibrated simulation boxes of heptane confined in (a) CS1000a (b) polymer at 5:95 volume ratio (c) kerogen-IIC. Heptane is in green (C) and white (H). The matrix is in black (C), gray (H), red (O), purple (S) and yellow (N), with a surface mesh. The images are generated using OVITO.\cite{stukowski:MSMSE2009}} \label{fig:Matrices}
		\end{center}
	\end{figure*}	
	
	In a series of papers since 2017 \cite{singer:jmr2017,singer:jcp2018,singer:jcp2018b,singer:EF2018,asthagiri:jpcb2020,valiyaparambathu:jpcb2020,singer:jpcb2020}, we have used atomistic MD (molecular dynamics) simulations to clarify the fundamental physics underlying NMR relaxation in fluids entirely avoiding the assumptions in traditional theories and without using \emph{adjustable} parameters. We found that \textit{intra}molecular $^1$H-$^1$H dipole-dipole relaxation in the fast-motion (i.e., low viscosity) regime for bulk water and alkanes \cite{singer:jmr2017,singer:jcp2018,asthagiri:jpcb2020} is not a mono-exponential as predicted by BPP while predicting $T_{1,2}$ in agreement with measurements. The simulations have helped clarify the importance of molecular flexibility \cite{singer:jcp2018} and spin-rotation \cite{singer:jcp2018b} in the relaxation response and revealed the ability of simulations to capture relaxation in the presence of the paramagnetic Gd$^{3+}$  \cite{singer:pccp2021,pinheiro:pccp2022}. In particular, for $n$-heptane in a viscous polymer matrix \cite{singer:EF2018,valiyaparambathu:jpcb2020,singer:jpcb2020}, 
	simulations showed that the anomalous $T_1$ plateau at high viscosity emerges due to nano-confinement, and a feature that had been explained by invoking paramagnetism \cite{serve2021}.
	
	Here we study the $^1$H-$^1$H dipole-dipole relaxation of heptane in 
	realistic kerogen models. Modeling kerogen is an arduous task and researchers have taken quite different approaches. We focus on two models: CS1000a, a rigid carbon matrix that has a pore distribution similar to Marcellus kerogen \cite{jain:lang2006,falk:NC2015}, and kerogen-IIC, a flexible molecular model of kerogen that respects the chemical composition based on an oil-prone mature kerogen from marine shales \cite{Collell:2014,Ungerer:2015}. We also include our previous results of heptane-polymer mixes \cite{valiyaparambathu:jpcb2020}, where the viscous polymer acts as a model for immature kerogen. All simulations presented here are at 25$^{\circ}$C, and all measurements are at 30$^{\circ}$C. 
 
	\ctable[
	caption={Summary of simulation results for confined heptane.},
	label=tab:Tab1,
	pos=ht,
	captionskip=-1.5ex,
        doinside={\small},
	left
	]	
	{cccccc}
	{
	\tnote{Table includes percent volume of heptane $\phi_{\mu}$ in the matrix, tortuosity of the matrix ${\mathcal T} = D_{bulk}/D_{sim}$ derived from the  diffusion coefficient under confinement $D_{sim}$ and bulk $D_{bulk}$, average correlation times $\left<\tau\right>$ computed using Eq. \ref{eq:Avg_tau}, $T_1$ at $f_0 = $ 2.3 MHz using Eq. \ref{eq:T1}, and residual dipolar coupling $T_{2,RDC}$ using Eq. \ref{eq:RDCsubtract}. All quantities are derived without any models or free parameters.}
	}
	{
	\FL
    Confining & $\phi_{\mu}$ & ${\mathcal T}$ & $\left<\tau\right>$ &  $T_1$ & $T_{2,RDC}$ \NN  
	matrix	  & (vol\%)      & (--)	          &  (ps)               & (ms)   & (ms) \ML
	Bulk 		 &100  &1 		&1.8    &8600 & $\infty$ \NN
	CS1000a		 &58   &80		&870    &29   &0.085     \NN
	Polymer 5:95 &5    &320	    &1500   &20   &0.120  \NN
	Kerogen-IIC	 &15   &370	    &1900   &13   &0.085   \ML
	}

	The three different confining matrices are illustrated in Figure \ref{fig:Matrices}. Listed in Table \ref{tab:Tab1} are the volume percent of heptane in the matrix, traditionally called ``micro'' porosity $\phi_{\mu}$ \cite{singer:fuel2020}. The heptane in CS1000a is assumed to be fully saturated, and hence the porosity $\phi_{\mu} =$ 58 vol\% is the net free volume available in the CS1000a matrix\cite{jain:lang2006}. For CS1000a, we also assume each matrix carbon atom as a site for $^1$H dipole relaxation. The heptane-polymer mixture is prepared with 5:95 volume ratio, hence $\phi_{\mu} =$ 5 vol\% \cite{valiyaparambathu:jpcb2020}. The heptane porosity in kerogen-IIC $\phi_{\mu} =$ 15.2 vol\% is adjusted based on swelling ratio, and is consistent with the measured porosity $\phi_{\mu} =$ 13.6 vol\% of dissolved heptane in Kimmeridge kerogen \cite{singer:fuel2020}. See SI for more details about the simulation.

	
	
	\textbf{$T_1$ dispersion}: The $^1$H-$^1$H dipole-dipole $T_1$ relaxation dispersion is derived from simulations by computing the autocorrelation function $G(t)$ for fluctuating magnetic $^1$H-$^1$H dipole-dipole interactions in the system \cite{bloembergen:pr1948, torrey:pr1953, abragam:book, mcconnell:book, cowan:book}. For an $isotropic$ system, $G(t)$ is given as: 
	\begin{align}
		&G(t) = \frac{1}{4}  \left(\frac{\mu_0}{4\pi}\right)^2 \hbar^2 \gamma^4 I(I+1)\times \nonumber \\
		&\frac{1}{N} \! \sum\limits_{i \neq j}^{N} \! \left< \frac{(3\cos^{2}\!\theta_{ij}(t+t')-1)}{r_{ij}^3\!\left(t+t'\right)}  \frac{(3\cos^{2}\!\theta_{ij}(t')-1)}{r_{ij}^3(t')} \right>_{\!\! t'}
		\label{eq:Gt}
	\end{align}
	where $t$ is the lag time of the autocorrelation, $\mu_0$ is the vacuum permeability, $\hbar$ is the reduced Planck constant, $\gamma/2\pi( = 42.58$ MHz/T) is the nuclear gyro-magnetic ratio for $^1$H (spin $I=1/2$), $r_{ij}$ is the magnitude of the vector connecting the $(i,j)$ $^1$H-$^1$H dipole-pairs, $\theta_{ij}$ is the polar angle between $\vec{r}_{ij}$ and the external magnetic field $\vec{B_0}$, and $N$ is the number of $^1$H's associated with heptane in the simulation box. The autocorrelation function $G(t)$ can be computed using the positions of all the hydrogens from the simulation trajectory. 
	
	Note that we neglect the other harmonic terms for $G(t)$ \cite{cowan:book} in Eq. \ref{eq:Gt} since the system is ``isotropic" over the length scale of the simulation box. Specifically,
	the orientation of an adsorbed heptane in kerogen is random with respect to $\vec{B_0}$, therefore the ensemble average over all randomly oriented heptane molecules averages out any preferred orientation over the length scale of the simulation box. This was confirmed by changing the direction of $\vec{B_0}$ in the simulation, resulting in a $\pm$6\% variation in $T_1$, which is consistent with uncertainties reported in Ref.~\citenum{pinheiro:pccp2022}. We note however that this would not generally be the case if the simulation were comprised of heptanes adsorbed onto a single nano-cylinder, for example; in such cases, all harmonic terms in $G(t)$ would be required, or a ``powder average'' is required \cite{faux:pre2013}.
	
	Once we have the autocorrelation function $G(t)$, we need the spectral density function $J(\omega)$ to compute $T_1$: 
	\begin{equation}
		J(\omega) = 2\int_{0}^{\infty}G(t)\cos\left(\omega t\right) dt,
		\label{eq:FourierRTcos}
	\end{equation}
	for $G(t)$ in units of $1/\rm{s}^2$ \cite{mcconnell:book}. As we can see in Eq.~\ref{eq:FourierRTcos}, we ideally need the \emph{entire} $G(t)$ curve to compute $T_1$ dispersion. However, in the slow-motion regime, $G(t)$ does not completely decay to zero in simulation time scales. (We use a maximum lag time of $t_{max} = $ 3 ns in $G(t)$ for trajectory data of length 10~ns.) We circumvent the role of limited data at large $t$ by expanding $G(t)$ as
	\begin{equation}
		G(t) = \int_{0}^{\infty} P(\tau) \exp \! \left( -\frac{t}{\tau} \! \right) d\tau.
		\label{eq:GtPtau} \, 
	\end{equation}
	and determining the underlying probability distribution $P(\tau)$ in correlation times $\tau$ \cite{singer:jpcb2020,valiyaparambathu:jpcb2020}, where we interpret the peaks in $P(\tau)$ as dynamic molecular modes \cite{asthagiri:jpcb2020}. This inversion technique has also found recent success in high-$T_c$ superconductors \cite{singer:prb2020} and quantum spin liquids \cite{wang:nature2021}. 
	Notice that Eq.~\ref{eq:GtPtau} is a Fredholm integral of the first kind with the kernal function  $\exp ( -t/\tau)$. We solve this using Tikhonov regularization to obtain $P(\tau)$, from which the average correlation time is defined as:
	\begin{equation}
		\langle \tau \rangle = \frac{1}{G(0)}\int_0^{\infty} P(\tau) \tau d \tau,
		\label{eq:Avg_tau}
	\end{equation}
	Once we have the distribution $P(\tau)$, Eq.~\ref{eq:FourierRTcos} reduces to the following:
	\begin{equation}
		J(\omega) = 2\int_{0}^{\infty}\frac{\tau}{1+ (\omega \tau)^2} P(\tau) d\tau.
		\label{eq:FourierPtau}
	\end{equation}
	and  $T_1$ is given by \cite{mcconnell:book,cowan:book}:
	\begin{equation}
		\frac{1}{T_{1}} = J(\omega_0) + 4 J(2\omega_0),  \label{eq:T1}
	\end{equation}
	where $\omega_0 = 2\pi f_0 = \gamma B_0$ is the Larmor (i.e., resonance) frequency, and $B_0$ the applied magnetic field strength. In the fast-motion (i.e., low viscosity) regime defined as $\omega_0\! \left<\tau\right> \ll 1$, $T_1= T_2 \propto 1/\!\left<\tau\right>$ and there is $no$ dispersion in $T_1$. In the slow-motion (i.e., high viscosity) regime defined as $\omega_0\! \left<\tau\right> \gg 1$, $T_1 > T_2$ and there $is$ dispersion in $T_1$. For crude oils and polymers, the transition $\omega_0\! \left<\tau\right> \simeq 1$ occurs at a viscosity of $\eta \simeq 6.4 \times 10^3 \, {\rm cP}/f_0$ (at 25$^{\circ}$C) for $f_0$ in (MHz) units \cite{liu:fuel2023}. A similar concept exists for fluids under nano-confinement, where $\left<\tau\right>$ increases with the degree of confinement. See SI for simulation results including $G(t)$, $P(\tau)$, $T_2$ dispersion, and intra vs intermolecular relaxation.
	
	As listed in Table \ref{tab:Tab1}, heptane in kerogen-IIC yields the longest correlation time $\left<\tau\right>$, followed by heptane in polymer, then heptane confined in CS1000a. Longer $\left<\tau\right>$ implies increasing confinement, indicating that kerogen-IIC is the most confining matrix for heptane, followed by the polymer mixture, followed by CS1000a. This can also be deduced from $T_1$ at low-frequencies (e.g., $f_0 < $ 2.3 MHz) where the fast-motion regime relation $T_{1}\propto 1/\!\left<\tau\right>$ holds, implying that shorter $T_1$ at $f_0 = $ 2.3 MHz indicates increasing confinement (see Table \ref{tab:Tab1}). Interestingly, this trend is also reflected in the simulated diffusion coefficient $D_{sim}$ of the confined heptane, which we express in Table \ref{tab:Tab1} in terms of the dimensionless tortuosity ${\mathcal T}$ defined as $D_{bulk}/D_{sim} = {\mathcal T}$, where $D_{bulk}$ = 3.43$\times$10$^{-9}$ m$^2$/s is the diffusion coefficient for bulk heptane \cite{valiyaparambathu:jpcb2020}. Our simulations indicate that increasing confinement, which manifests itself as increasing $\left<\tau\right>$ and decreasing $T_1$ at $f_0 = $ 2.3 MHz, correlates with the tortuosity ${\mathcal T}$ of the confining medium.	
	
	\begin{figure}[h!]
		\centering
		\includegraphics{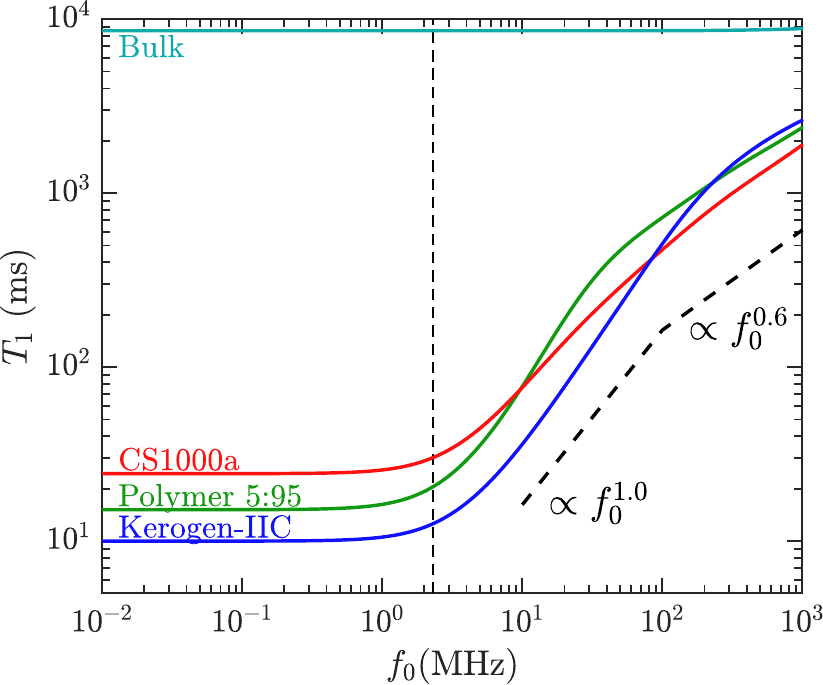}
		\caption{$T_1$ dispersion for heptane confined in the two models of kerogen studied. The bulk heptane, and heptane confined in poly(isobutene), are also plotted for comparison. The dashed line indicates $f_0=$ 2.3 MHz. The functional forms $T_1 \propto f_0^{\beta}$ are also plotted. }\label{fig:T1_dispersion}
	\end{figure}
	
	\begin{figure}[!h]
		\centering
		\includegraphics{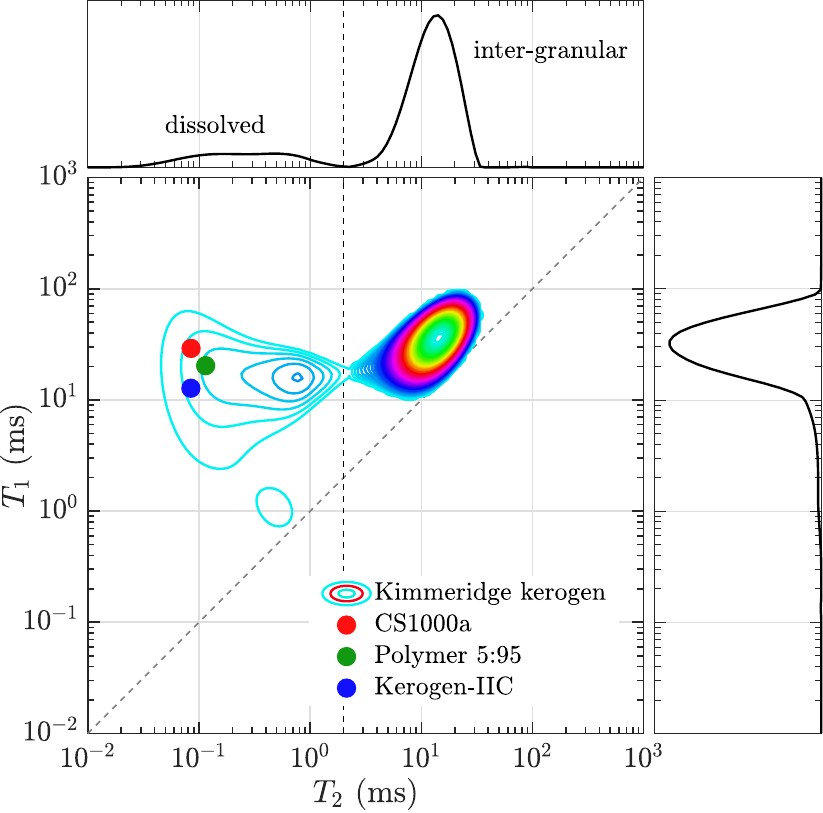}
		\caption{2D $T_1$-$T_2$ map measurement of heptane-saturated kerogen isolates from the Kimmeridge Shale at $f_0$ = 2.3 MHz \cite{singer:fuel2020}, $T_1$ ($T_2$) projections at the top (right) subplots (respectively), 1-1 diagonal line $T_1 = T_2$, and vertical cutoff line separating dissolved heptane ($T_2 < $ 2 ms)  from inter-granular heptane. Also shown are simulation results from Table \ref{tab:Tab1}, \textit{without any models or free parameters.}}
		\label{fig:2D_map}
	\end{figure}		
	
	Fig.~\ref{fig:T1_dispersion} shows the computed $T_1$ dispersion (i.e., frequency dependence), where $T_1$ shows significant dispersion at higher frequencies ($f_0 > $10 MHz), as expected from our previous work~\cite{valiyaparambathu:jpcb2020}. Specifically, heptane in polymer 5:95 and kerogen-IIC shows approximate $T_1 \propto f_0^{1}$ behavior between 10 $< f_0 < $ 100 MHz followed by shallower $T_1 \propto f_0^{0.6}$ behavior above $ f_0 > $ 100 MHz, consistent with viscous fluids \cite{singer:jpcb2020}. Meanwhile heptane in CS1000a shows $T_1 \propto f_0^{0.6}$ behavior at $ f_0 > $ 10 MHz and above. The universal form $T_1 \propto f_0^{0.6}$ above $ f_0 > $ 100 MHz for all three systems shows the lack of dependence of $T_1$ on the level of confinement for heptane, analogous to $T_1$ plateauing at high $\eta/T$ for viscous fluids. Furthermore, the functional form $T_1 \propto f_0^{0.6}$ is consistent with measurements of heptane confined in organic-rich chalk \cite{liu:fuel2023}. In general, a power-law dependence $T_1 \propto f_0^{\beta}$ suggests relaxation by RMTD (reorientations mediated by translational displacements) \cite{gizatullin2022}. Further investigations are underway to see whether the dynamic molecular modes $P(\tau)$ can give insights into the RMTD mechanism and the surface topology.

\begin{figure}[!h]
		\centering
		\includegraphics{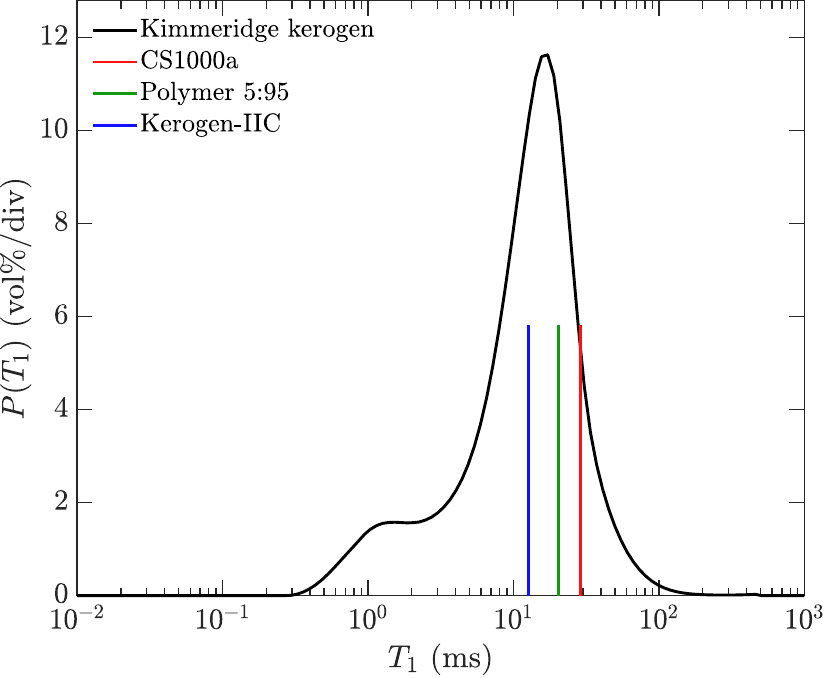}
		\caption{$T_1$ distribution for heptane-saturated Kimmeridge kerogen, taken from the $T_1$ projection of Figure \ref{fig:2D_map} for dissolved heptane signal alone ($T_2 < $ 2 ms). Also shown are simulation results from Table \ref{tab:Tab1}.}
		\label{fig:T1_dist}
	\end{figure}

	Fig.~\ref{fig:2D_map} shows the simulated $T_1$ values at $f_0 = $ 2.3 MHz from Table \ref{tab:Tab1} compared to the 2D $T_1$-$T_2$ measurement (also at $f_0 = $ 2.3 MHz) of heptane confined in kerogen isolates from Kimmeridge Shale taken from Ref.~\citenum{singer:fuel2020,singer:petro2016,chen:petro2017}, which is a Type II immature kerogen. See SI for more details about the experiment. The simulations give consistent values for $T_1$ compared to the measured signal for dissolved heptane ($T_2 <$ 2 ms). The projection of the measured $T_1$ distribution for dissolved heptane is shown Fig. \ref{fig:T1_dist} alongside the simulation results from Table \ref{tab:Tab1}. The kerogen-IIC matrix gives the closest match with the log-mean of the measured $T_1$ distribution ($T_{1LM} \simeq$ 11 ms) for dissolved heptane, \textit{without any models or free parameters.} We note that the realistic kerogen model used here provides greater surface area and surface roughness than smooth nano-slits and cylinders \cite{Mutisya:jpcc2017,Amaro-Estrada:Lang2022}. This results in much larger confinement effects in kerogen where the low-frequency $T_1$ is $\sim$3 orders of magnitude shorter than bulk, compared to a factor $\sim$2 for smooth nano-surfaces.
	
	\textbf{$T_2$ residual dipolar coupling}: 
	In the fast-motion (i.e., low-frequency) regime, the simulations would normally predict that $T_2 \simeq T_1$ lies on the 1-1 diagonal of the 2D map, which is clearly not consistent with $T_2$ measurements for dissolved heptane in Figure \ref{fig:2D_map}. Similar inadequacies in $T_2$ have also been noted in other nano-confined systems \cite{chen:jpcb2020,chen:acs2022}. 
	Upon reflection, we can infer that in contrast to the bulk, there is anisotropy in the system originating from strong adsorption onto the kerogen surface, leading to additional dephasing (i.e.. transverse relaxation) in $T_2$. This
    residual dipolar coupling (RDC) has been hypothesized to 
    play an important role in organic-rich shale \cite{washburn:jmr2017,washburn:cmr2014}. The RDC
    is given by the following expression\cite{washburn:jmr2017}:
	\begin{align}
		\Delta\omega_{RDC}  &= \frac{1}{2}\frac{\mu_0}{4 \pi} \gamma^2 \hbar \times \nonumber \\
		&\cfrac{1}{N_i} \sum_{i=0}^{N_i} \left|\sum_{j\neq i}^{N_j} \left<\frac{3\cos^{2} \theta_{ij}(t) - 1}{r_{ij}^3 (t)} \right>_{\!\!t} \right|
		\label{eq:RDCterm}
	\end{align}
	in units of (rad/s), where $r_{ij}$ is the distance between $^1$H dipoles $i$ and $j$, and $\theta_{ij}$ is the angle the vector connecting the two dipoles make with the external magnetic field. We also compute the RDC for bulk heptane $\Delta\omega_{RDC}^{\rm bulk}$, then take the difference between confined heptane and bulk to determine $T_{2,RDC}$ as such
    \begin{align}
    	\frac{1}{T_{2,RDC}} = \Delta\omega_{RDC} - \Delta\omega_{RDC}^{\rm bulk}
    	\label{eq:RDCsubtract}
    \end{align}
    SI provides details about the RDC calculations.
	
	As listed in Table \ref{tab:Tab1}, we find a comparable $T_{2,RDC} \simeq$ 0.085 ms for CS1000a and kerogen-IIC, while $T_{2,RDC} \simeq$ 0.120 ms for polymer 5:95 is longer, which may be due to the more flexible nature of polymer 5:95 compared to kerogen-IIC or CS1000a. The $T_{2,RDC} \simeq$ 0.085 ms for kerogen-IIC implies a RDC of $\Delta f_{RDC} = 1/(2\pi T_{2,RDC}) \simeq$ 1900 Hz, which is $\sim$1 order of magnitude larger than previous reports of heptane confined in 6 nm diameter silica nano-tubes $\Delta f_{RDC} \simeq$ 140 Hz \cite{valiullin:pre2006}. The larger $\Delta f_{RDC}$ for kerogen-IIC is because heptane is completely ``adsorbed'' in the kerogen, i.e., the nano-pores are of order the size of the heptane molecule, resulting in larger confinement effects than the 6 nm silica nano-tubes.
	
	The distribution of $T_{2}$ relaxation times often found in measurements of porous media are due to the distribution in pore-sizes in the sample. In the present case, the relaxation induced by the pore, termed as ``surface relaxation" time, is equal to $T_{2,RDC} \simeq$ 0.085 ms for kerogen-IIC. The total relaxation rate is given by 
	the sum of the two relaxation rates, with the adsorbed heptane relaxing at $T_{2,RDC}$ (surface), and the pore fluid heptane relaxing at $T_{2,B}$ (bulk), both of which are in fast exchange with each other, i.e.\ in the fast diffusion regime \cite{brownstein:pr1979}. Under such conditions the total measured relaxation rate is given by the following expression \cite{brownstein:pr1979}: 
	\begin{equation}
		\frac{1}{T_{2}} = \frac{Sh}{V} \frac{1}{T_{2,RDC} + \tau_m} + \left(1 - \frac{Sh}{V}\right) \frac{1}{T_{2,B}}, \label{eq:SV}
	\end{equation}
	where $S$ is the surface area of the nano-pore, $V$ is the volume of the nano-pore, $h$ = 0.42 nm \cite{mao:jpcb2001} is the thickness of the adsorbed layer of heptane, and $\tau_m$ is the residence time of heptane on the kerogen surface. The term $Sh/V$ is essentially the fraction of adsorbed heptane on the surface. Given that $T_{2,RDC} \gg \tau_m$ and that the bulk relaxation term is negligible, we can derive a pore-size distribution from the measured $T_2$ distribution as such:
	\begin{equation}	
		\frac{1}{T_{2}} \simeq \frac{h}{T_{2,RDC}}  \frac{S}{V} \simeq \rho_2 \frac{4}{d}, \label{eq:cylinder}
	\end{equation}
	where we assume the nano-pores to be cylindrical of diameter $d$, i.e., $S/V = 4/d$. The surface relaxivity parameter $\rho_2 = h/T_{2,RDC} =$ 4.9 nm/ms (or 4.9 $\mu$m/s, equivalently) is ``calibrated'' from the simulated $T_{2,RDC}$. This calibration procedure assumes that in the dissolved state, like in the MD simulation cases, all heptanes are completely ``adsorbed", hence $Sh/V = 4h/d = 1$ in the simulations. The simulated $\rho_2$ is consistent with previous reports in organic-rich shale calibrated from scanning electron microscopy (SEM) images \cite{rylander:spe2013}.

	\begin{figure}[h]
		\centering
		\includegraphics{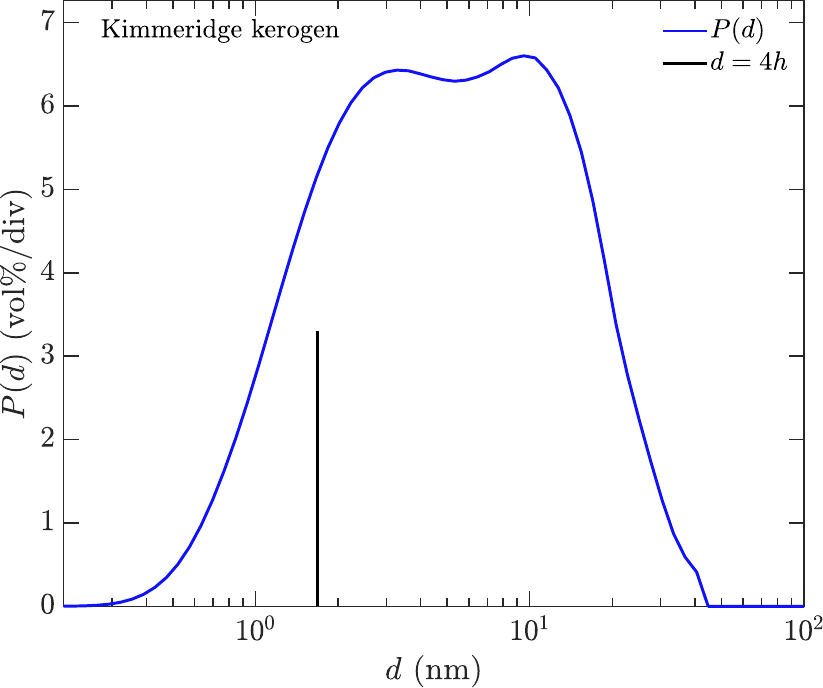}
		\caption{Pore-size distribution $P(d)$ of dissolved heptane in the Kimmeridge kerogen from the $T_2$ projection (below $T_2< $ 2 ms) in Figure \ref{fig:2D_map}, where the pore diameter $d = 4 \rho_2 T_2$. The surface relaxivity $\rho_2 = h/T_{2,RDC} $ is calibrated from simulations, with the calibration point $d=4h$ (i.e., $T_2 = T_{2,RDC}$) shown.}
		\label{fig:poresizedist}
	\end{figure}
	
	Fig.~\ref{fig:poresizedist} shows the pore-size distribution $P(d)$ of dissolved heptane, obtained using Eq. \ref{eq:cylinder} with $\rho_2$ calibrated from simulations. It is remarkable that $P(d)$ has the potential to provide an accurate pore-size distribution of the organic nano-pores in kerogen from NMR relaxation measurements and MD simulations alone, without any external experimental data to calibrate the NMR pore-size distribution such as SEM images or N$_2$ gas adsorption (BET) isotherms. 
	
	A similar derivation of pore-size distribution from $T_1$ is not presented due to strong cross-relaxation effects in $T_1$ \cite{liu:fuel2023}, which are not present in $T_2$. Cross-relaxation has a tendency to average out and narrow the $T_1$ distribution from all $^1$H in contact with each other through spin diffusion (a.k.a. magnetization transfer) \cite{gerig2011,washburn:cmr2014,singer:jcp2018,chen:jpcb2020,chen:acs2022}. In the present case, the $^1$H bearing molecules in contact with each other include inter-granular heptane, dissolved heptane, bitumen, and kerogen. Note that cross-relaxation is independent of the fact that the bitumen and kerogen are not detectable here due to their fast solid-like $T_{2G} \simeq 0.01$ ms relaxation and instrumental limitations at low frequencies $f_0$ = 2.3 MHz (typically, higher frequency relaxometers with shorter dead-times are required to detect solid-like signals \cite{zamiri2021,liu:fuel2023}). 
	Cross-relaxation in the present case is evidenced by the narrower $T_1$ distribution compared to $T_2$ in Fig. \ref{fig:2D_map}, which makes the pore-size distribution $P(d)$ more accurate from $T_2$ than from $T_1$. On the other hand, the dispersion in $T_1$ allows for a wealth of information about the dynamic molecular modes in $P(\tau)$, making the $T_1$-$T_2$ correlation a very powerful combination for fluids under nano-confinement.

	In summary, we use MD simulations of heptane confined in realistic kerogen models to compute the NMR $^1$H-$^1$H dipole-dipole relaxation including $T_1$ relaxation dispersion and $T_2$ residual dipolar coupling. We find that heptane $T_1$ relaxation is strongly enhanced by confinement in organic nano-pores, with $T_1$ at low-frequencies reduced by $\sim$3 orders of magnitude compared to bulk. The simulated $T_1$ values agree with the measurements of heptane in kerogen from the Kimmeridge Shale, \textit{without any models or free parameters.} We find that the $T_2$ relaxation is dictated by residual dipolar coupling, with $T_2$ reduced by $\sim$5 orders of magnitude compared to bulk. We use the simulated $T_{2}$ to calibrate the surface relaxivity $\rho_2$, which is used to convert the measured $T_{2}$ distribution of dissolved heptane into the pore-size distribution $P(d)$ of the organic nano-pores in kerogen, \textit{without additional experimental data}. Importantly, the measurements of heptane confined in kerogen can be explained by $^1$H-$^1$H dipole-dipole relaxation, without invoking the physics of paramagnetism. Furthermore, our simulation techniques (detailed in the SI) are relevant to $^1$H-$^1$H dipole-dipole relaxation for all nano-confined systems.
	
	\section{Supporting Information} 
	{The Supporting Information includes (1) simulation methodology, (2) additional simulation results, specifically the simulated autocorrelation functions, $T_2$ dispersion, and separation of intra- and inter-molecular contributions, (3) residual dipolar coupling methodology, and (4) experimental details.}
	
	\section{Acknowledgements}
	We thank Chevron Energy Technology Company, the Rice University Consortium on Processes in Porous Media, and the American Chemical Society Petroleum Research Fund (No. ACS PRF  58859-ND6) for financial support. 
	We gratefully acknowledge the National Energy Research Scientific Computing Center, which is supported by the Office of Science of the U.S. Department of Energy (No.\ DE-AC02-05CH11231) and the Texas Advanced Computing Center (TACC) at The University of Texas at Austin for high-performance computer time and support. 
	Research at Oak Ridge National Laboratory is supported under contract DE-AC05-00OR22725 from the U.S. Department of Energy to UT-Battelle, LLC. This research used resources of National Energy Research Scientific Computing Center, which is supported by the Office of Science of the U.S. Department of Energy under Contract \# DE-AC02-05CH11231. ORNL Pub.\ ID.\ 188552. 
	
	
	\providecommand{\latin}[1]{#1}
\makeatletter
\providecommand{\doi}
  {\begingroup\let\do\@makeother\dospecials
  \catcode`\{=1 \catcode`\}=2 \doi@aux}
\providecommand{\doi@aux}[1]{\endgroup\texttt{#1}}
\makeatother
\providecommand*\mcitethebibliography{\thebibliography}
\csname @ifundefined\endcsname{endmcitethebibliography}
  {\let\endmcitethebibliography\endthebibliography}{}

\end{document}


\newpage

\setcounter{page}{1}
\makeatletter
\renewcommand{\thepage}{ S\@arabic\c@page }

\setcounter{section}{0}
\makeatletter
\renewcommand{\thesection}{ S.\@arabic\c@section }

\setcounter{figure}{0}
\makeatletter
\renewcommand{\thefigure}{S\@arabic\c@figure}

\setcounter{table}{0}
\makeatletter
\renewcommand{\thetable}{S\@arabic\c@table}

\setcounter{equation}{0}
\makeatletter
\renewcommand{\theequation}{S.\@arabic\c@equation}

\advance\cftsecnumwidth 0.6em\relax
\advance\cftsubsecnumwidth 0.6em\relax
\advance\cftsubsubsecnumwidth 0.6em\relax

\tableofcontents

\clearpage

\section{Simulation methodology}
	
	\subsection{Heptane in mesoporous carbon matrix}

	\subsubsection{Matrix structure}
	
	We used the CS1000a mesoporous carbon matrix as an idealization of the kerogen system\cite{jain:lang2006}. The matrix itself is rigid, but interestingly the pore distribution is similar to that in Marcellus Kerogen and gives a good description of diffusivity
	for alkanes in kerogen under reservoir conditions\cite{falk:NC2015}. The carbon atoms forming the wall 
	are modeled after graphene carbon atom with
	12-6 Lennard-Jones $\sigma$ = 3.550 \AA~and $\epsilon$ = 35.222 kJ/mol.

	\begin{figure}[b!]
		\begin{center}
			\includegraphics[width=1\columnwidth]{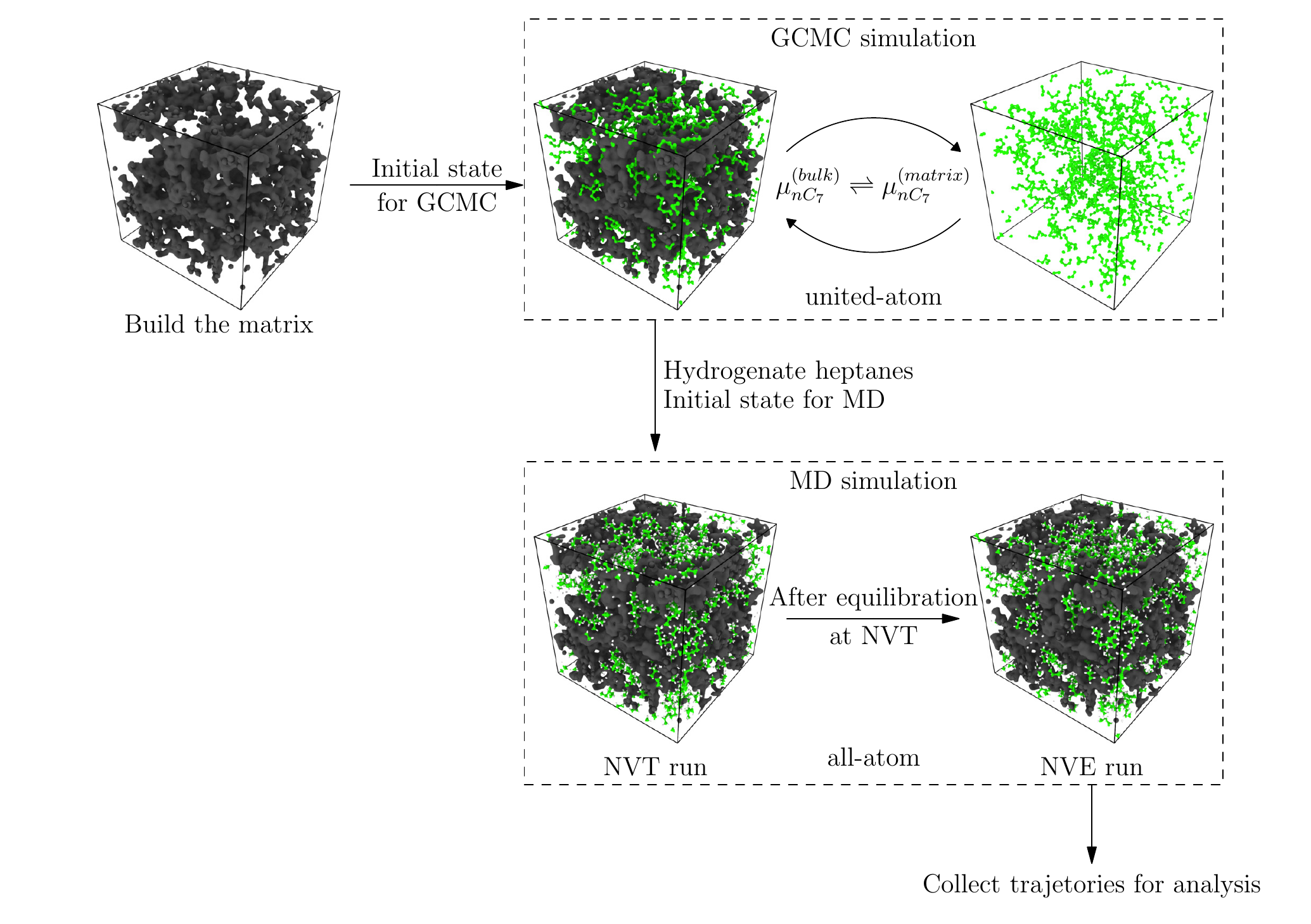}
			\caption{Schematic of the simulation details for heptane in cs1000a carbon matrix.}\label{fg:csgcmc}
		\end{center}
	\end{figure}

	\subsubsection{Grand canonical Monte Carlo simulations}
	
	The CS1000a matrix was populated with heptane using the grand canonical Monte Carlo simulation technique using Cassandra\cite{cassandra}. The system was equilibrated with a bulk heptane system corresponding to 298 K and 1 atm. The configurational bias technique was used to improve the sampling\cite{shah:jcp2011}. We simplified the MC simulation by opting for a united atom force field in this step. The heptane was modeled using TraPPE force field\cite{martin:jpcb1998}. The system was equilibrated for 1$\times$10$^5$ steps, and then the data was collected over a subsequent run of 3$\times$10$^6$ steps. The number of molecules in the system was observed to be stable at 249 molecules. We use the last configuration for further exploration using molecular dynamics. 
	
	\subsubsection{Molecular Dynamics simulations}
	
	We use NAMD\cite{NAMD, phillips:JCC2005} to perform the molecular dynamics simulations. In this phase, we explicitly describe hydrogens, which were added using the PSFGEN tool to the previously saved united atom heptane configuration. In other words, we assume each matrix carbon atom as a site for $^1$H dipole relaxation. The heptanes are described using the CHARMM General Force Field \cite{cgenff}. The CS1000a matrix was modeled using  12-6 Lennard-Jones sphere with $\sigma$ = 3.550 \AA~and $\epsilon$ = 35.222 kJ/mol, exactly as the GCMC simulations. The Lennard-Jones interactions were terminated at $14$ \AA~with smooth switching from $13$ \AA. The Coulombic interactions were cutoff at $13$ \AA, and we use the particle mesh Ewald procedure to describe long-range electrostatic interactions. We minimized the system to remove any strain using the conjugate gradient method. The equations of motion were propagated using the RESPA multistep algorithm with a base time step of $1$ fs. The system was equilibrated at a constant temperature of $298.15$ K for $1$ ns. The temperature was controlled by reassigning velocities (obtained from a Maxwell-Boltzmann distribution) every $250$ fs. The thermostat was then turned off and the production was carried out for $5$ ns. 
	
	\subsection{Heptane in Kerogen II-C Model}
	
	\subsubsection{Matrix structure and MD simulation details}
	
	The molecular model for different types of kerogen at different maturities was developed by Ungerer {\it et al.} \cite{Ungerer:2015}. In this study, we use Kerogen II-C model, which is an oil prone mature kerogen from marine shales. The H/C ratio for this kerogen is 0.90 and the O/C ratio is 0.05. We use LAMMPS\cite{LAMMPS, plimpton:JCoP1995} to perform the molecular dynamics simulations. The PCFF+ force field\cite{yiannourakou:OGST2013} (an extension of PCFF \cite{sun:JACS1994}) was used to model the system. The parameters were obtained from MedeA$\textsuperscript{\textregistered}$ software environment\cite{francelanord:euro2014} and Collel {\it et al.} \cite{Collell:2014}. The Lennard-Jones and Coulombic interactions were terminated  at 15 \AA. We use the particle-particle particle-mesh procedure to describe the long range electrostatic interactions. The equations of motions were propogated using the velocity Verlet alogirthm with a timestep of 1~fs. For equilibration runs, the temperature and pressures were controlled using a three chained Nos\'e-Hoover thermostat \cite{nose:mp1984,hoover:pra1985} and a Nos\'e-Hoover barostat \cite{hoover:pra1986} respectively.

	\begin{figure}[ht!]
		\begin{center}
			\includegraphics[width=0.95\columnwidth]{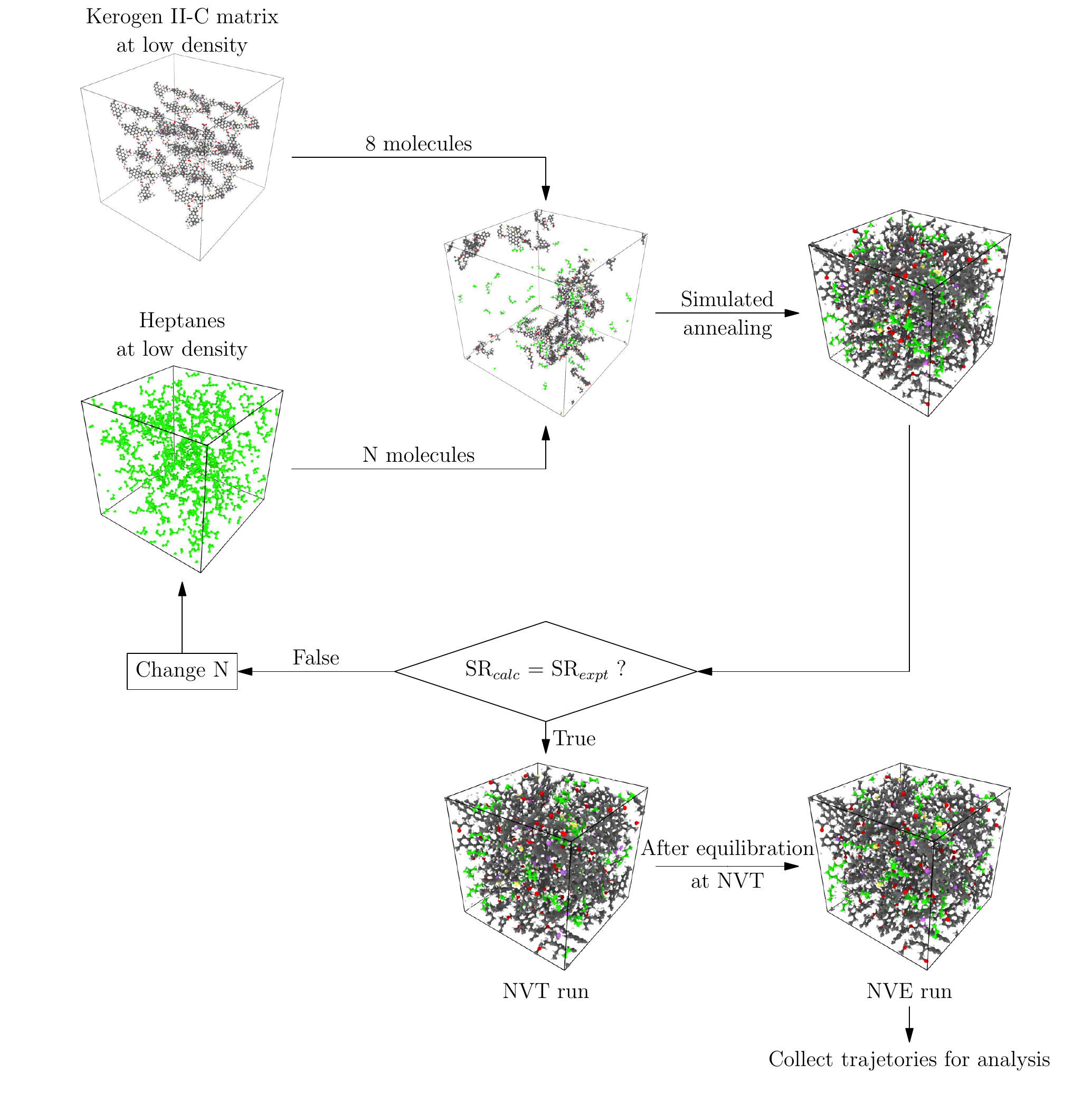}
			\caption[Schematic of the simulation details for heptane in kerogen II-C]{Schematic of the simulation details for heptane in kerogen II-C. Note that, unlike heptane in CS1000a system where we used a GCMC simulations to create the initial structure, we utilize experimental estimate of dissolved heptane to estimate the initial structure.  }\label{fg:keriic}
		\end{center}
	\end{figure}
	
	\subsubsection{Creating the initial structure}
	
	In order to populate the matrix at approximately the right thermodynamic conditions, we use the simulated annealing technique. We create 8 copies of kerogen molecule and $N$ copies of heptane, and pack them at a very low density (0.1 g/cm$^3$) using in-built tools in LAMMPS. We next compress the mixture to obtain the target density. We turn off the long-range electrostatics and use a LJ and Coulombic cutoffs of 9 \AA, since only short-range interactions are critical in molecular packing. We then perform an NVT simulation at 1000 K for 100 ps. This step ensures reasonable mixing of the system and removes any biases created while making the initial structure. This is followed by NPT simulations of 200 ps to compress the system to 100 atm. We then slowly reduce the temperatures from 900 K to 300 K  in 4 steps of 200 ps. We then restore the long-range electrostatics, and change the LJ and Coulombic cutoffs to 15 \AA. Then we equilibrated the system for 1 ns at the required state of 298.15 K and 1 atm pressure.
	
	\begin{figure}[ht!]
		\begin{center}
			\includegraphics[width=0.49\columnwidth]{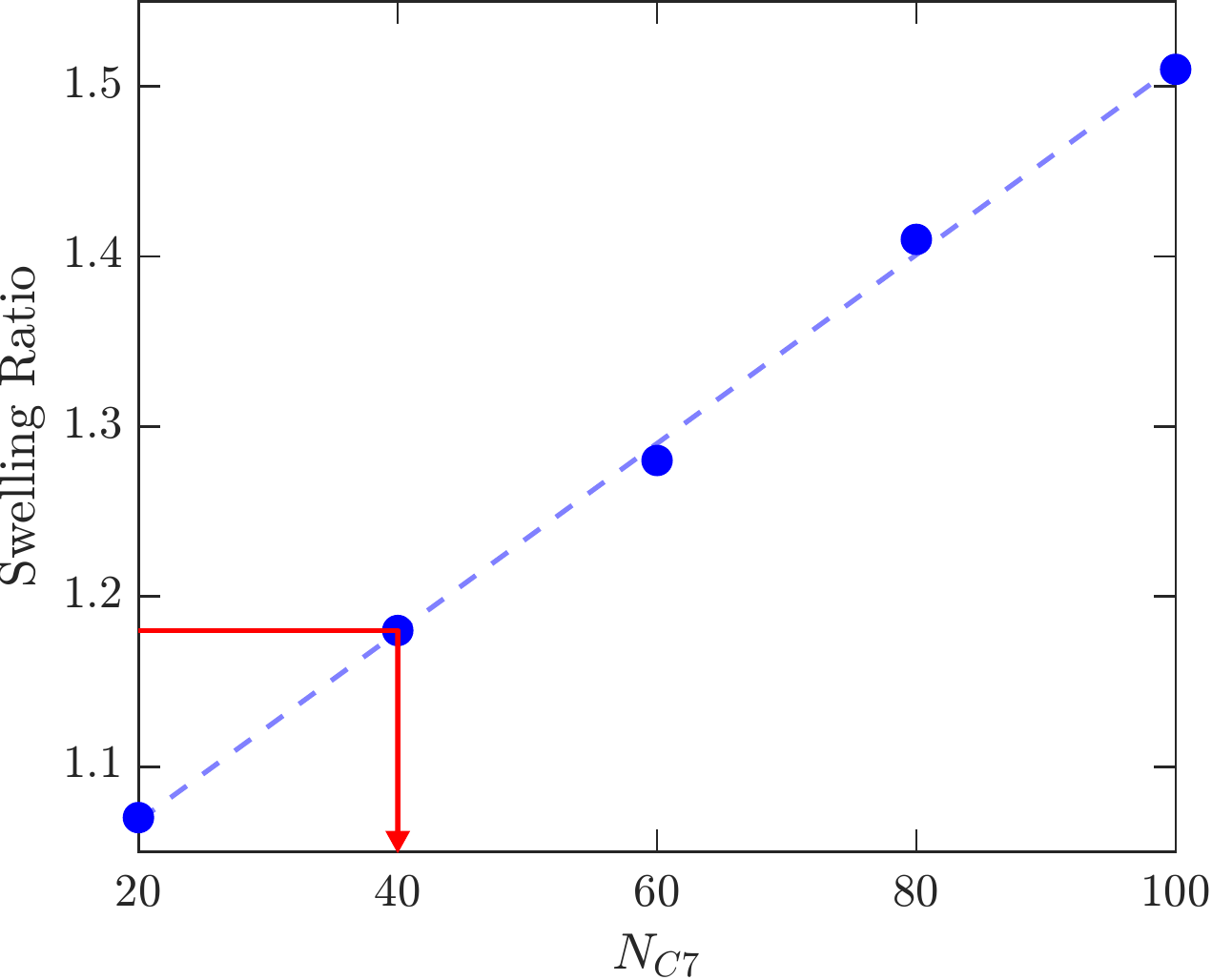}
			\caption[Swelling ratio for different number of heptane molecule]{Swelling ratio for different number of heptane molecule. The estimate for heptane swelling ratio in kerogen II-C due to dissolved heptane is 1.18\cite{singer:fuel2020}.}\label{fg:swelling}
		\end{center}
	\end{figure}
	
	The number of heptanes was changed from 20 molecules to 100 molecules. We found that 40 molecules of heptane gave us the right swelling ratio for the given kerogen model. Note that the swelling ratio $SR = 1.18$ can also be expressed as a volume percent heptane 
	\begin{equation}
			\phi_{\mu} = 100\left(1-\frac{1}{SR}\right) \quad {\rm (vol\%)}, \label{eq:SR}
	\end{equation}
	yielding $\phi_{\mu} = $ 15.2 vol\%, which is consistent with the measurement of $\phi_{\mu} = $ 13.6 vol\%. The system with 40 heptane molecules in 8 molecules of Kerogen-IIC was simulated for production for 10 ns at constant energy. 

    \subsection{Diffusion coefficient}
    The diffusion coefficient was computed using Einstein relation
    \begin{equation}
        D_{sim} = \lim_{t \to \infty} \frac{1}{6}\frac{\delta \langle \Delta r^2 \rangle}{\delta t}
        \label{eq:Dsim}
    \end{equation}
    where $\delta \langle \Delta r^2 \rangle$ is the mean-square displacement of the center-of-mass of the heptane as a function of evolution time $t$. We sample all the heptanes in each of the matrix to compute the diffusion coefficient, and evaluated for diffusion evoultion time of 1 ns.  Note that we do not apply the finite system size corrections here as an estimate of viscosity is required \cite{yeh:jpcb2004}, which is ill-defined for a rigid matrix system. Our previous study on heptane-polymer matrix indicate that at the correction is $\sim$9\% for heptane in bulk $D_{bulk}$, and decreases to $\sim$2\% for the polymer 5:95 system.  

	\begin{table}[!ht]
		\centering
		\hspace*{-0.5cm}\begin{tabular}{cccc} 
			\hline\hline
			Confining& $D_{sim}$ & $\mathcal{T}$  \\ 
			matrix	& (m$^2$s$^{-1}$/$10^{-11}$)  & (--) \\ 
			\hline
			Bulk 		 &343   &1      \\
			CS1000a		 &4.27   &80       \\
			Polymer 5:95 &1.05   &320     \\
			Kerogen-IIC	 &0.93   &370     \\
			\hline\hline
		\end{tabular}
		\caption{Self-diffusion coefficient of heptanes using Eq. \ref{eq:Dsim}, along with corresponding tortuosity values $\mathcal{T} = D_{bulk}/D_{sim}$.}
		\label{tab:TabDt}
	\end{table}	
	The tortuosity ${\mathcal T}$ defined as ${\mathcal T} = D_{bulk}/D_{sim} $ assumes that heptane is in the tortuosity limit, which is valid for dissolved heptane.
\clearpage	
\section{Additional simulation results}
	
	\subsection{Autocorrelation of dipole-dipole relaxation}
\begin{figure}[h!]
	    \begin{center}
	    	\includegraphics[width=2.5in]{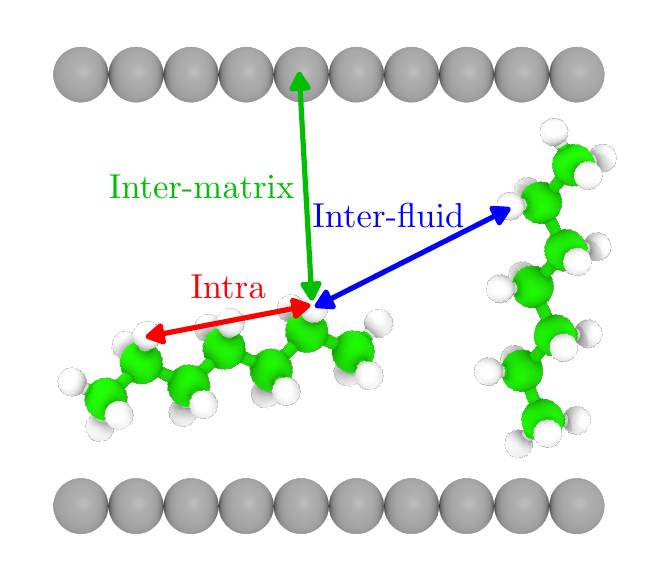}
	    	\caption{Partitioning into \textit{intra}molecular relaxation $G_R(t)$, and \textit{inter}molecular relaxation $G_T(t)$. Note that \textit{inter}molecular relaxation consists of both other heptanes and the matrix. }\label{fg:interintrawall}
    	\end{center}
    \end{figure}

	The autocorrelation $G(t)$ can be applied to all $^1$H's taken together or grouped according to whether the two protons are on the same molecule, leading to \textit{intra}molecular relaxation, or on separate molecules, leading to \textit{inter}molecular relaxation, see Fig. \ref{fg:interintrawall}. Such a partitioning is valid if there is no cross coupling between the two relaxation mechanisms, as is the case here. 
    We indicate \textit{intra}molecular relaxation by the subscript $R$ and \textit{inter}molecular relaxation by the subscript $T$. We follow Eq.~2 to 6 to obtain $T_{1,R}$ and $T_{1,T}$. The total relaxation is then given as
    \begin{equation}
	    \frac{1}{T_{1}} = \frac{1}{T_{1R}} + \frac{1}{T_{1T}}. \label{eq:T1T2total}
    \end{equation}
    
\begin{figure}[ht!]
		\begin{center}
			\includegraphics[width=0.49\textwidth]{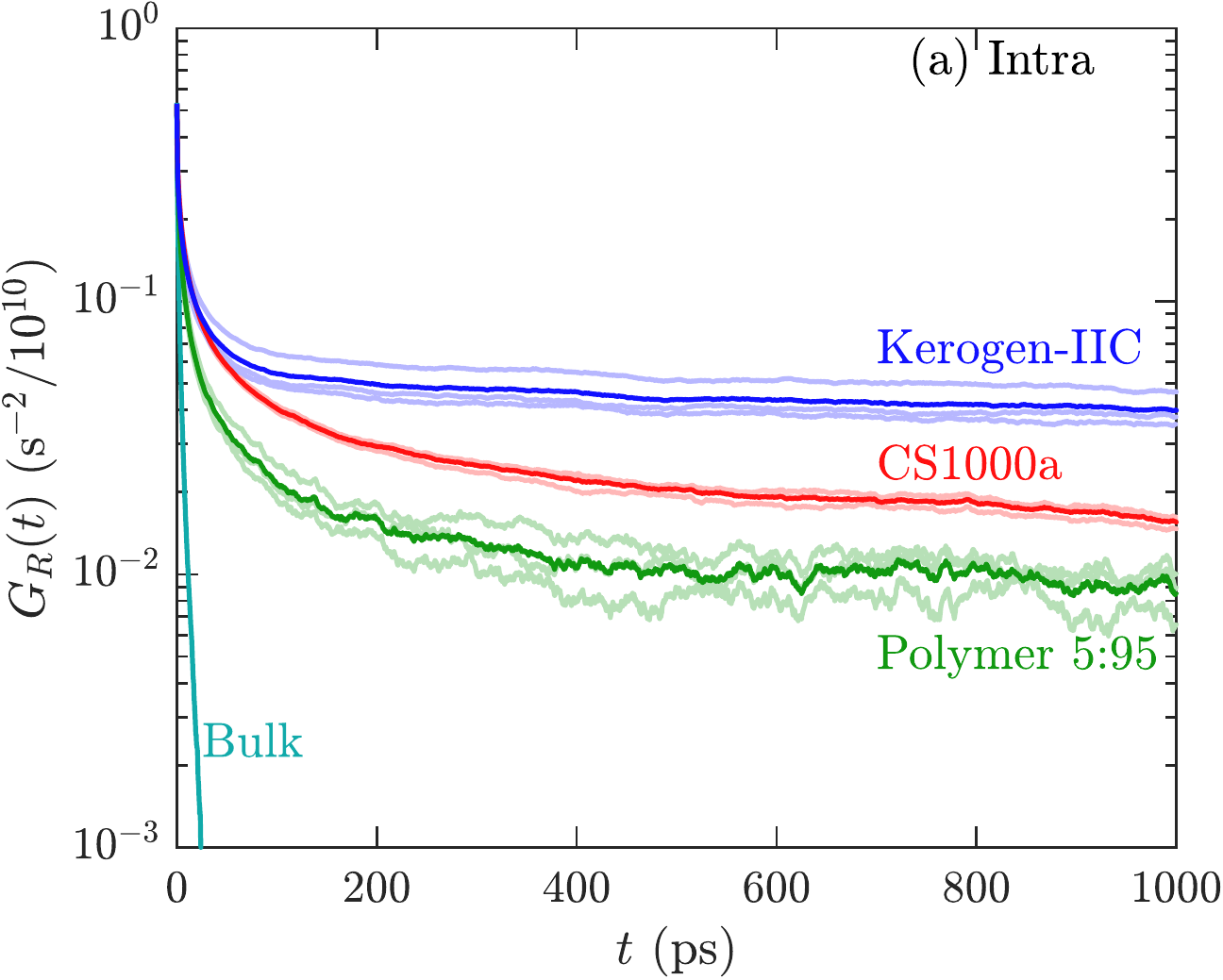}
			\includegraphics[width=0.49\textwidth]{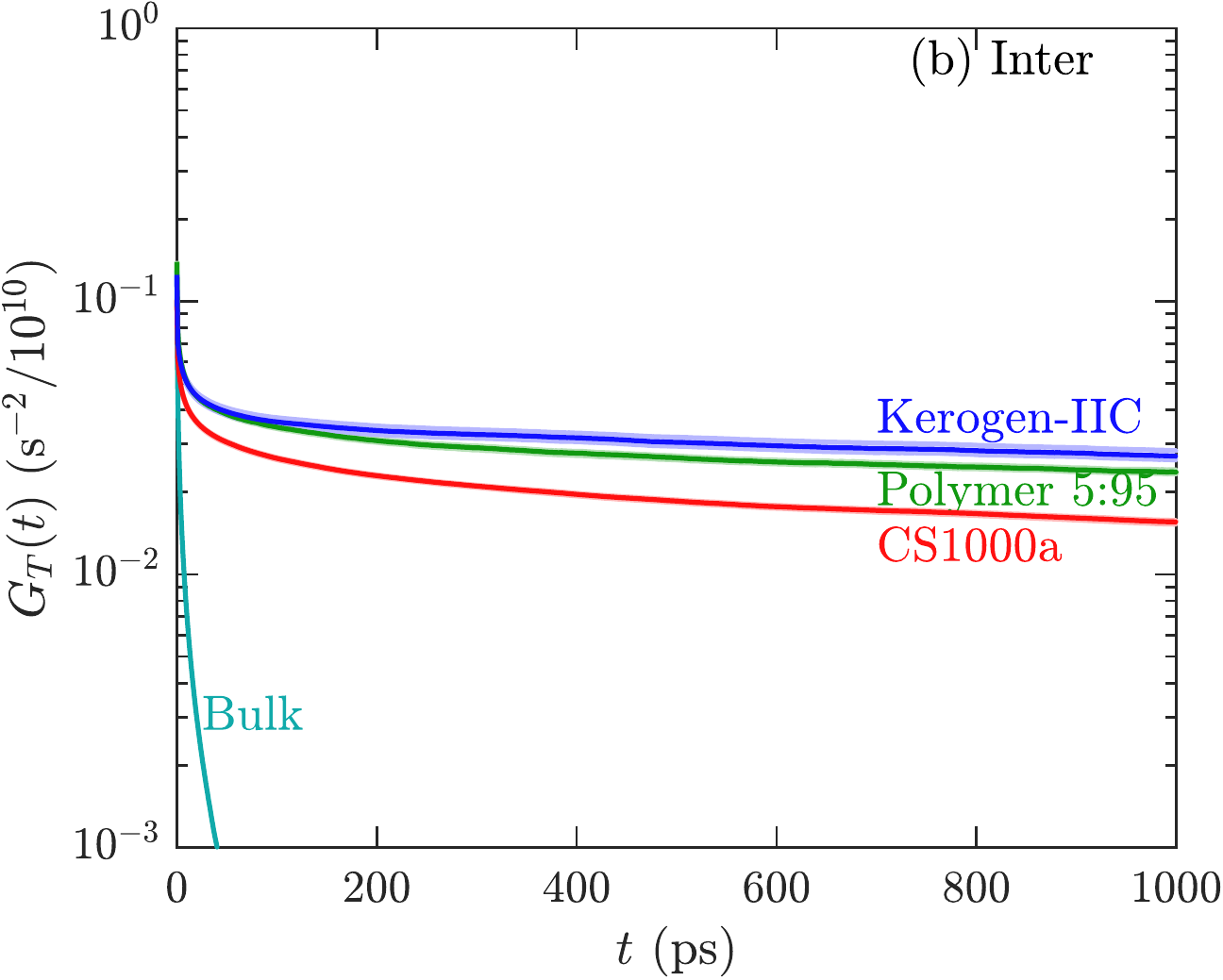}
			\caption{(a) \textit{Intra}molecular autocorrelation $G_R(t)$ and (b) \textit{inter}molecular autocorrelation $G_T(t)$ for heptane in CS1000a, polymer 5:95, and kerogen-IIC. The darker shade indicates the average $G_{RT}(t)$ ,and the lighter shades indicates the autocorrelation curve for different directions. The bulk $G_{RT}(t)$ is plotted for comparison.}\label{fg:Gtinterintra}
		\end{center}
	\end{figure}

	The \textit{intra}molecular and \textit{inter}molecular autocorrelation functions $G_{R,T}(t)$ were computed using Eq.~1 from the last $6.5536$ ns of the simulations, and upto lag time of $t_{max} = 3$ ns is utilized. It should be noted that for CS1000a the matrix is made of carbon atoms alone, hence the matrix effect was computed assuming each carbon atom as a site for a $^1$H dipole. This is included in the \textit{inter}molecular autocorrelation calculations. The direction of the magnetic field $\vec{B_0}$ was chosen to be the three primary axes $\vec{x}$, $\vec{y}$, and $\vec{z}$, and the mean $G_{R,T}(t)$ was taken for further analysis. The computed autocorrelations are shown in Fig.~\ref{fg:Gtinterintra}.
    
\begin{figure}[ht!]
		\begin{center}
			\includegraphics[width=0.49\textwidth]{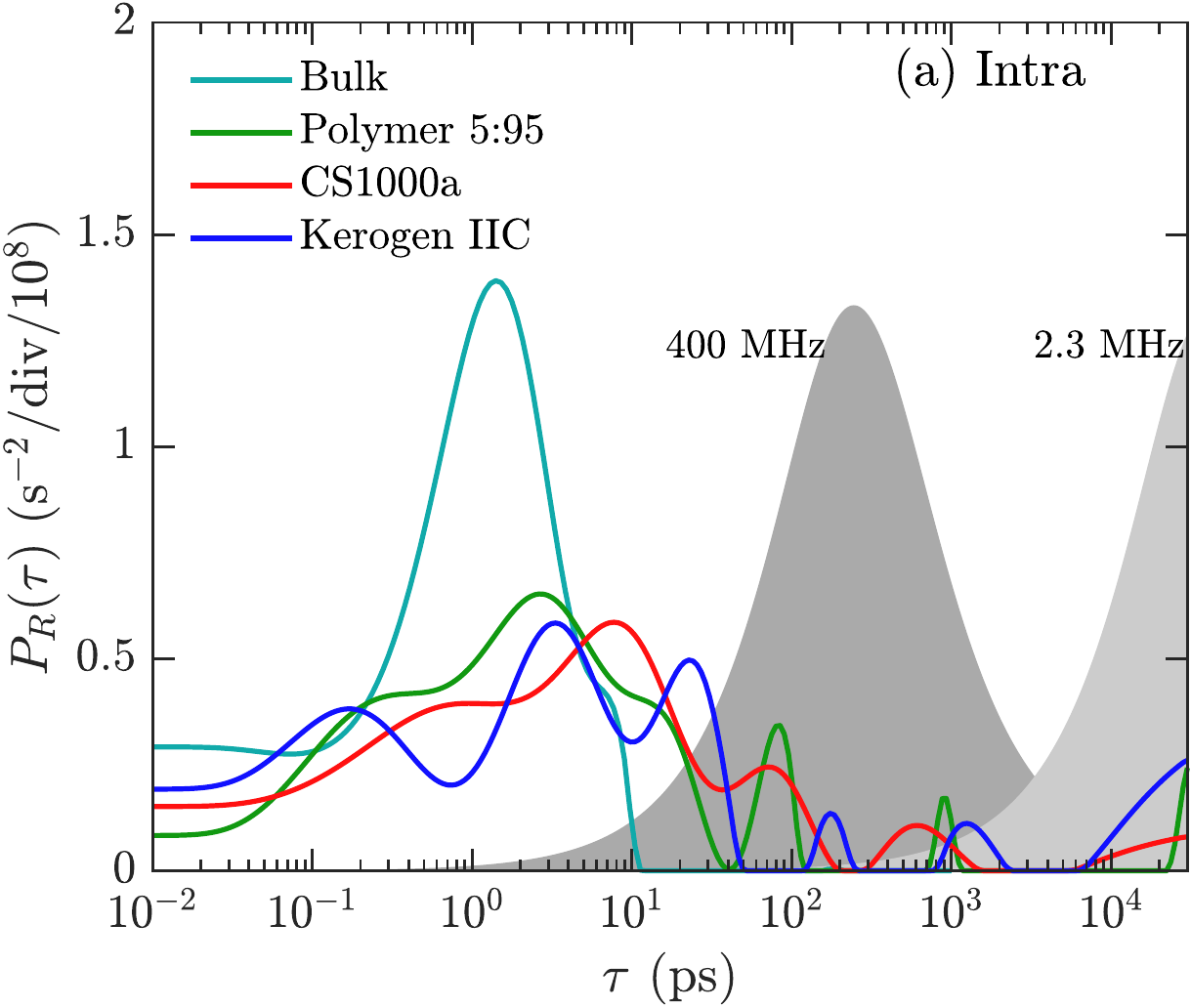}
			\includegraphics[width=0.49\textwidth]{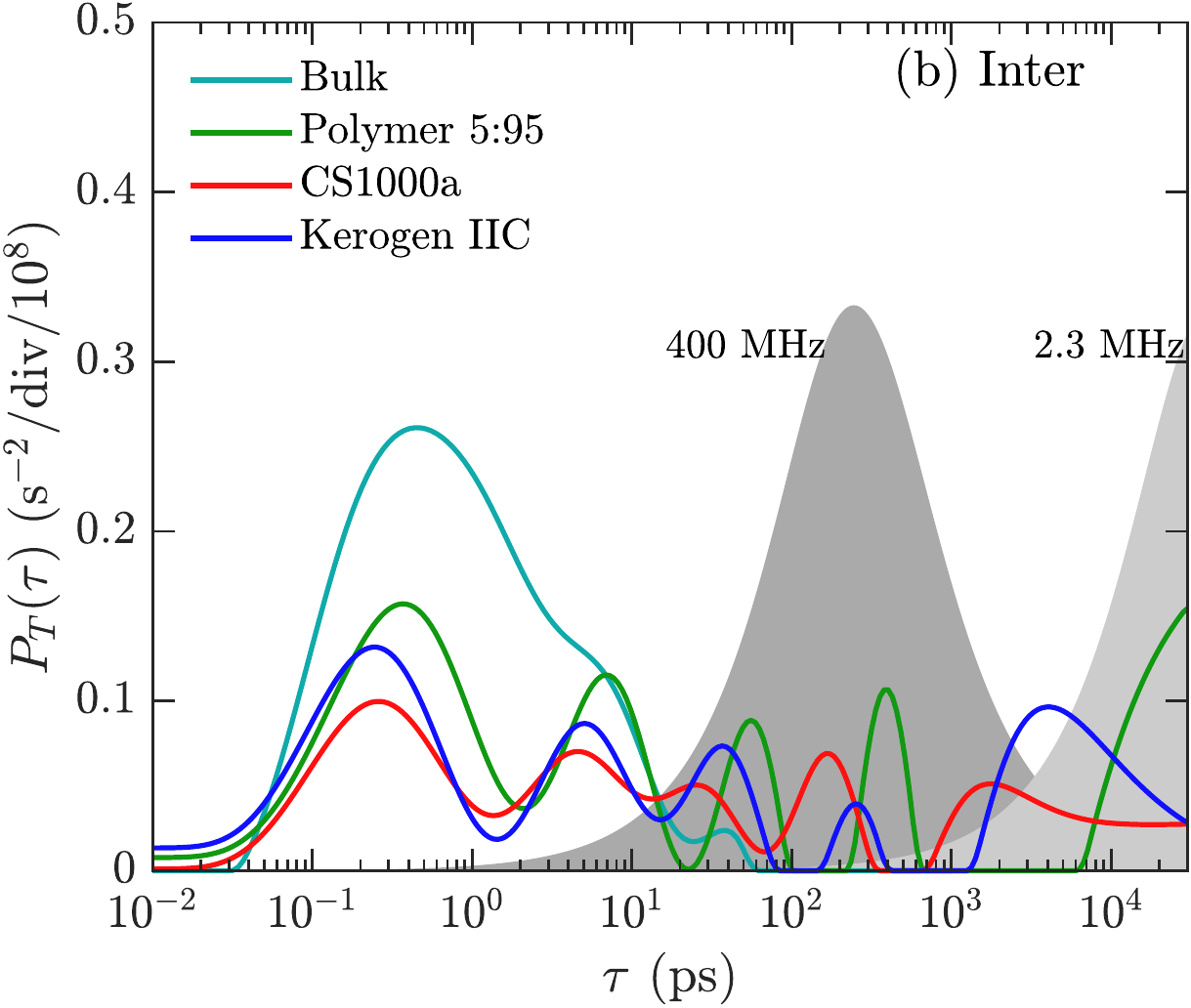}
			\caption{Probability distribution of correlation times for (a) \textit{intra}molecular autocorrelation $P_R(\tau)$ and (b) \textit{inter}molecular autocorrelation $P_T(\tau)$ for heptane in CS1000a, polymer 5:95, and kerogen-IIC. The bulk $P_{RT}(\tau)$ is plotted for comparison. The gray area plot shows the $T_1$ filter $\mathcal{L}$ at frequencies 2.3 MHz and 400 MHz respectively.}\label{fg:Ptauinterintra}
		\end{center}
	\end{figure}

    The distribution of correlation times $P_{RT}(\tau)$ was evaluated using Eq.~3, with $\tau_{min} = 10^{-2}$ ps, $\tau_{max} = 10 \,t_{max} = 30 $ ns, and fixed regularization parameter $\alpha = 1$\cite{singer:jpcb2020,valiyaparambathu:jpcb2020}. The computed distributions are shown in Fig.~\ref{fg:Ptauinterintra}.
    
    We termed the ``peaks" observed in the $P(\tau)$ distribution as molecular modes, as it is empirically observed to be functions of molecular structure and its environment. Note that the two shaded areas are the $T_1$ filters at 2.3 MHz and 400 MHz respectively. $T_1$ filter signifies the ``weight" of a given correlation time towards $T_1$ at a given frequency. The $T_1$ filter $\mathcal{L}$ for given frequency $\omega_0$ is given by
    \begin{equation}
    \mathcal{L}(\omega_0) \propto \Bigg [ \frac{2 \tau}{1 + (\omega_0 \tau)^2} + \frac{8 \tau}{1 + (2 \omega_0 \tau)^2} \Bigg ].
    \end{equation} 
    
    We can see that, as the frequency increases, the lower molecular modes are more important. This also highlights that simulations capture NMR relaxation at higher frequencies more accurately. This is an important uncertainity in our results of $T_1$ values at 2.3 MHz. 

    \clearpage
	\subsection{Intra vs Intermolecular relaxation}
    
\begin{figure}[h!]
	\centering
	\includegraphics[width=3.2in]{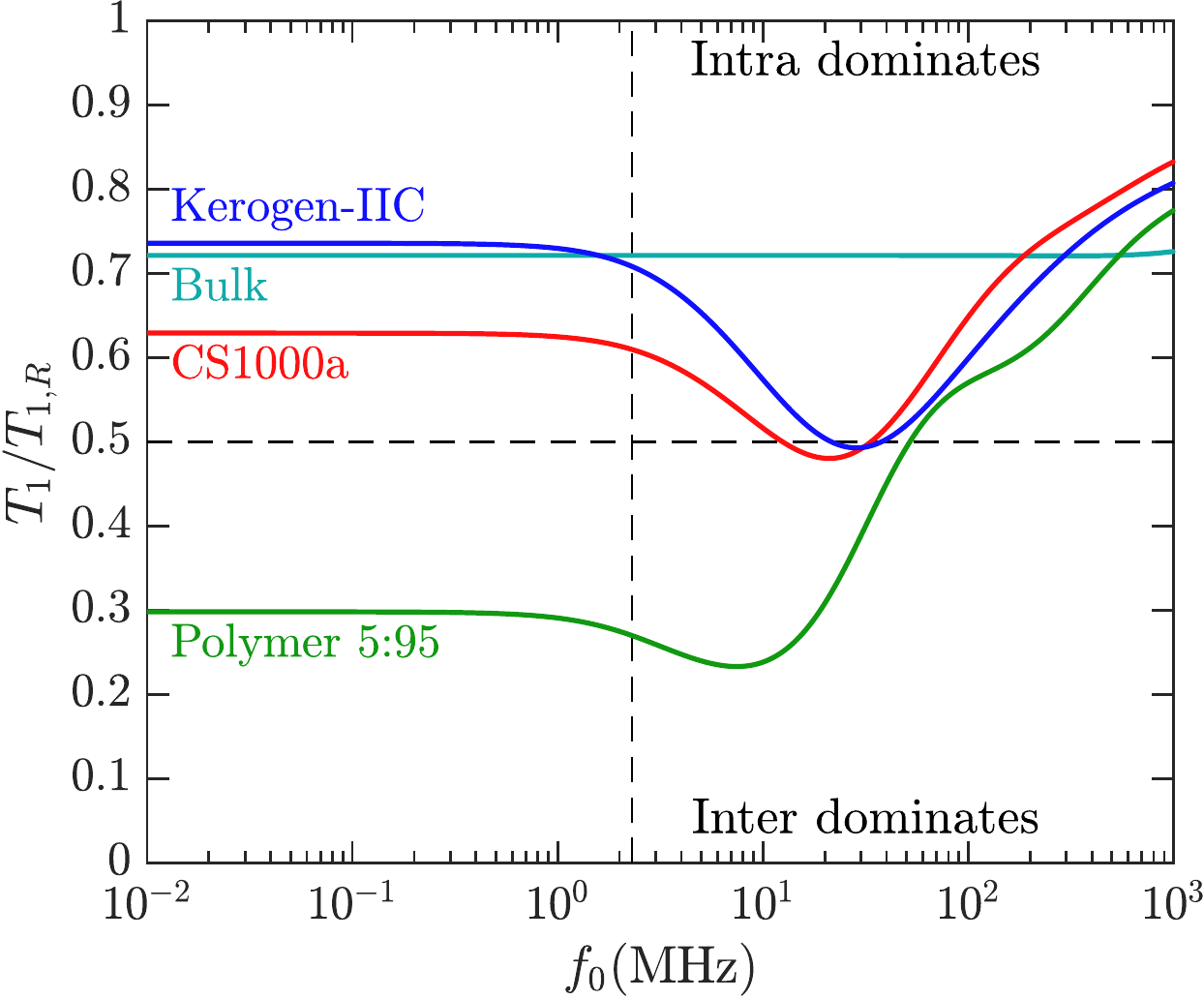}
	\caption{The ratio of \textit{intra}molecular relaxation rate to the total relaxation rate, where $T_1/T_{1R} > $ 0.5 indicates that \textit{intra}molecular dominates, while $T_1/T_{1R} < $ 0.5 indicates that \textit{inter}molecular dominates. The vertical dashed line indicates the frequency of 2.3 MHz.}\label{fg:T1ratio}
    \end{figure}

    The effect of the matrix can be elucidated by looking at the individual components of relaxation. Fig.~\ref{fg:T1ratio} shows the ratio of $T_{1,R}$ relaxation in total relaxation $T_1$. 

    Fig.~\ref{fg:T1ratio} indicates that at low-frequencies ($f_0 \ll $ 2.3 MHz), the int\textit{ra}molecular relaxation $T_{1R}$ plays a more significant role in heptane confined in kerogen-IIC and CS1000a compared to polymer. This is due to higher rigidity of the kerogen-IIC and CS1000a matrices compared to the polymer matrix, which considerably affects the int\textit{ra}molecular motion of the heptane molecule. At higher frequencies ($f_0 \gg $ 100 MHz), the int\textit{er}molecular contribution diminishes faster than the int\textit{ra}molecular. This can be explained by the $T_1$ filter on $P_{RT}(\tau)$ distribution, with greater importance to shorter molecular modes at higher frequencies\cite{valiyaparambathu:jpcb2020} and for $\tau<10^3$ ps, $P_{R}(\tau)>P_{T}(\tau)$. 

\clearpage
\subsection{$T_2$ dispersion} 

$T_2$ for $^1$H-$^1$H dipole-dipole relaxation is given by \cite{mcconnell:book,cowan:book}:
\begin{equation}
	\frac{1}{T_{2}} = \frac{3}{2} J(0) + \frac{5}{2} J(\omega_0) + J(2\omega_0), \label{eq:T2}
\end{equation}
which predicts a mild increase of 10/3 in $T_2$ with increasing $\omega_0\!\left<\tau\right>$. However, all experimental data indicates that the factor 10/3 increase in $T_2$ is not observed in viscous fluids, and that $T_2$ remains frequency independent \cite{singer:jpcb2020}.

\begin{figure}[ht!]
	\centering
	\includegraphics[width=0.49\textwidth]{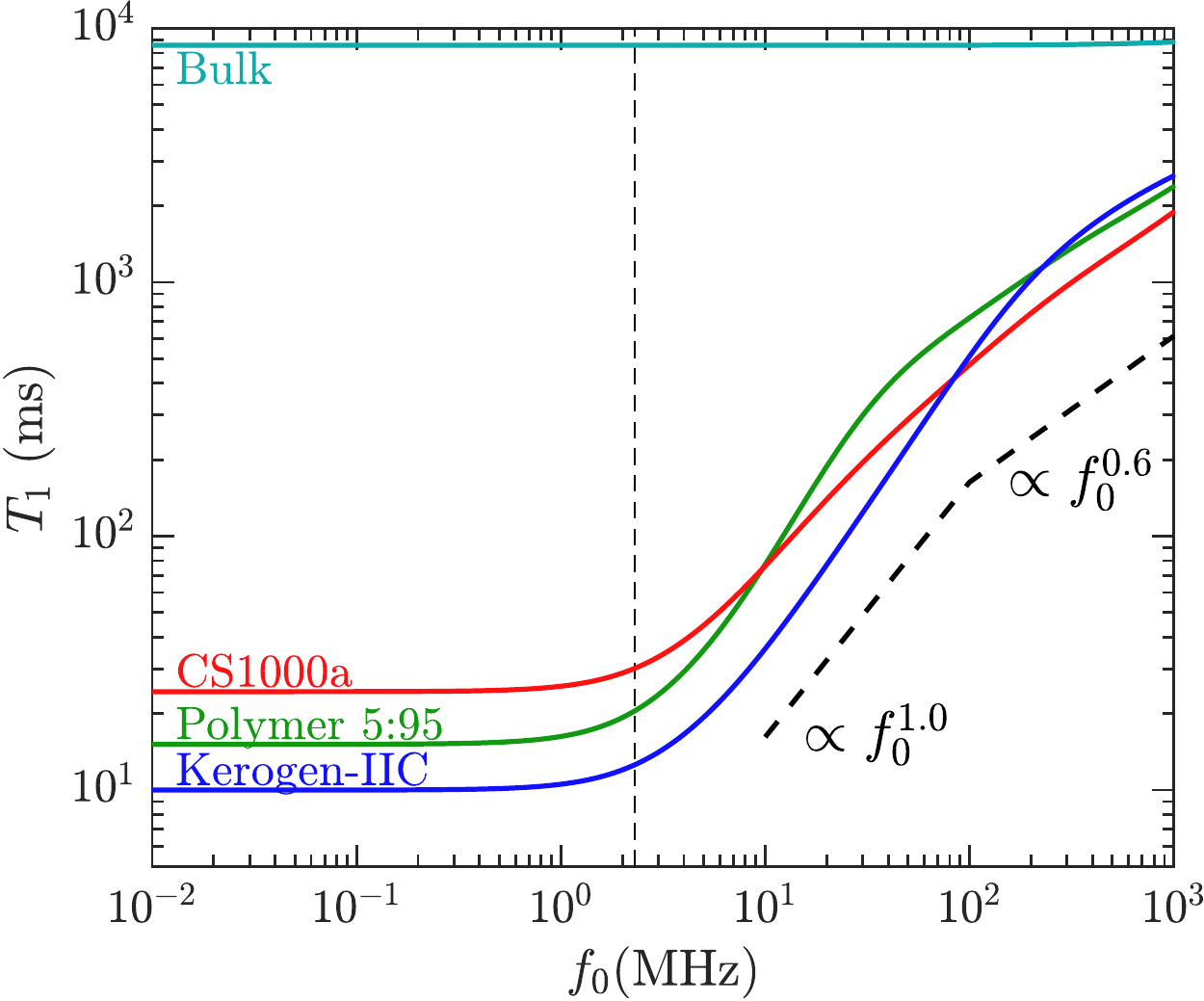}  
	\caption{NMR relaxation times $T_2$ for heptane in the two models of kerogen studied. The bulk heptane and heptane confined in poly(isobutene) are plotted for comparison.}\label{fg:kero:T2}
\end{figure}

\begin{figure}[!h]
	\centering
	\includegraphics[width=0.49\textwidth]{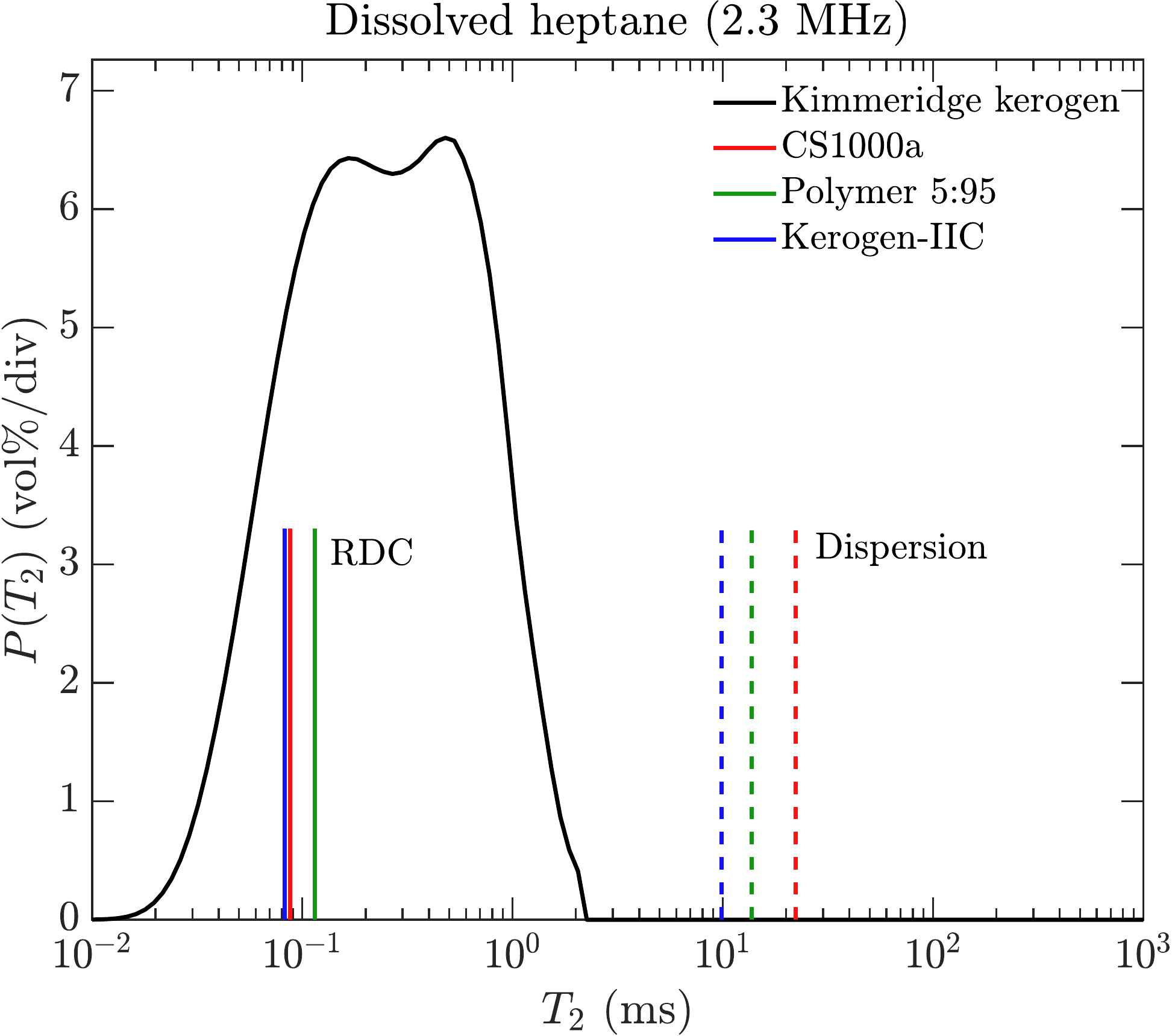}
	\caption{$T_2$ distribution for heptane-saturated kerogen isolates from the Kimmeridge Shale at $f_0$ = 2.3 MHz, taken from the $T_2$ projection of Figure 3 for dissolved heptane signal alone ($T_2 < $ 2 ms). Also shown are simulation results according to RDC from Eq. \ref{eq:RDCterm} (solid lines) and $T_2$ dispersion from Eq. \ref{eq:T2} (dashed lines), \textit{without any models or free parameters.}}
	\label{fig:T2_dist}
\end{figure}

The total transverse relaxation $T_{2,Total}$ is given by the sum of rates of Eq. \ref{eq:T2} and \ref{eq:RDCterm} as such:
    \begin{equation}
	    \frac{1}{T_{2,Total}} = \frac{1}{T_{2}} + \frac{1}{T_{2,RDC}}. \label{eq:T2total}
    \end{equation}
For confined heptane, the $T_2$ term is neglected since $T_2$ is $\sim$2 orders of magnitude longer than $T_{2,RDC}$, i.e., $T_{2,Total} \simeq T_{2,RDC}$. A comparison of the ``$T_2$ dispersion" term versus $T_{2,RDC}$ is shown in Figure \ref{fig:T2_dist}.
	
\clearpage
\section{Residual dipolar coupling methodology}	

The residual dipolar coupling $\Delta\omega_{RDC}$ \cite{washburn:jmr2017} is calculated using Eq.~7 (main text) and reproduced here below
	\begin{align}
		\Delta\omega_{RDC} &= \frac{1}{2}\frac{\mu_0}{4 \pi} \gamma^2 \hbar  \cfrac{1}{N_i} \sum_{i=0}^{N_i} \left|\sum_{j\neq i}^{N_j} \left<\frac{3\cos^{2} \theta_{ij}(t) - 1}{r_{ij}^3 (t)} \right>_{\!\!t} \right|
		\label{eq:RDCterm}
	\end{align}
in (rad/s) units, using all the $^1$H in the system and the last 6.5536 ns of the simulation. 

Since we are evaluating an anisotropic effect of the system from a periodic cubic simulation box, which is inherently anisotropic due to the cubic shape\cite{allentildesley:book} and absence of angular momentum conservation\cite{hoover:book,kuzkin:zamm2015}. Note that it is also well known that pair correlations evince periodicity artifacts which become
pronounced near the vertices and edges of the cubic simulation box \cite{mandell,pratthaan1,pratthaan2}. Hence, we need to exercise caution while evaluating Eq.~7.

The following steps were taken to remove simulation artifacts. The inner sum is only evaluated for hydrogens $j$ with $r_{ij}<15$~\AA~from the hydrogen $i$. The RDC was also evaluated for three primary axes $\vec{x}$, $\vec{y}$, and $\vec{z}$, and the mean RDC was used for further computations. For CS1000a, we once again assume each matrix carbon atom as a site for $^1$H dipole. The simulation box is still not expected to preserve isotropy, so we compute the RDC for bulk heptane $\Delta\omega_{RDC}^{\rm bulk}$, which serves as a baseline given bulk heptane is isotropic. Hence, the difference between confined heptane and bulk is used to determine $T_{2,RDC}$ as such
\begin{align}
	\frac{1}{T_{2,RDC}} = \Delta\omega_{RDC} - \Delta\omega_{RDC}^{\rm bulk}
	\label{eq:RDCsubtract}
\end{align}
resulting in the following values:
	\begin{table}[!ht]
		\centering
		\hspace*{-0.5cm}\begin{tabular}{cccc} 
			\hline\hline
			Confining& $\Delta\omega_{RDC}$ & $\Delta\omega_{RDC} - \Delta\omega_{RDC}^{\rm bulk}$ & $T_{2,RDC}$  \\ 
			matrix	& (k-rad/s) & (k-rad/s) & (ms) \\ 
			\hline
			Bulk 		 &1.55   &0     &$\infty$ \\
			CS1000a		 &13.3   &11.8  &0.085     \\
			Polymer 5:95 &10.3   &8.70   &0.120  \\
			Kerogen-IIC	 &13.4   &11.8  &0.085   \\
			\hline\hline
		\end{tabular}
		\caption{List of $\Delta\omega_{RDC}$ values using Eq. \ref{eq:RDCterm}, along with $T_{2,RDC}$ using Eq. \ref{eq:RDCsubtract}}
		\label{tab:TabRDC}
	\end{table}

	\clearpage
	\section{Experimental details}
	
	The Kimmeridge Shale originated from the coast of Dorset in the United Kingdom, and is the same as in Refs. \citenum{singer:petro2016},\citenum{chen:petro2017}, and\citenum{singer:fuel2020}. LECO and RockEval measurements found a total organic carbon of $TOC \simeq 10.1$ wt\% (or $\simeq 17.6 $ vol\% equivalently), a hydrogen index of HI $\simeq$ 684 mg\_HC/g\_orgC, an oxygen index of OI $\simeq$ 12 mg\_CO$_{(2)}$/g\_orgC, and a $T_{max}$ $\simeq$ 424 $^{\circ}$C, indicating Type II immature kerogen, or specifically Type II-A (immature) according to Ungerer {\it et al.} classification \cite{Ungerer:2015}. This is more immature than the simulated kerogen, which is Type II-C (middle-end of oil window).
	
\begin{figure}[!ht]
		\begin{center}
			\includegraphics[width=0.5\textwidth]{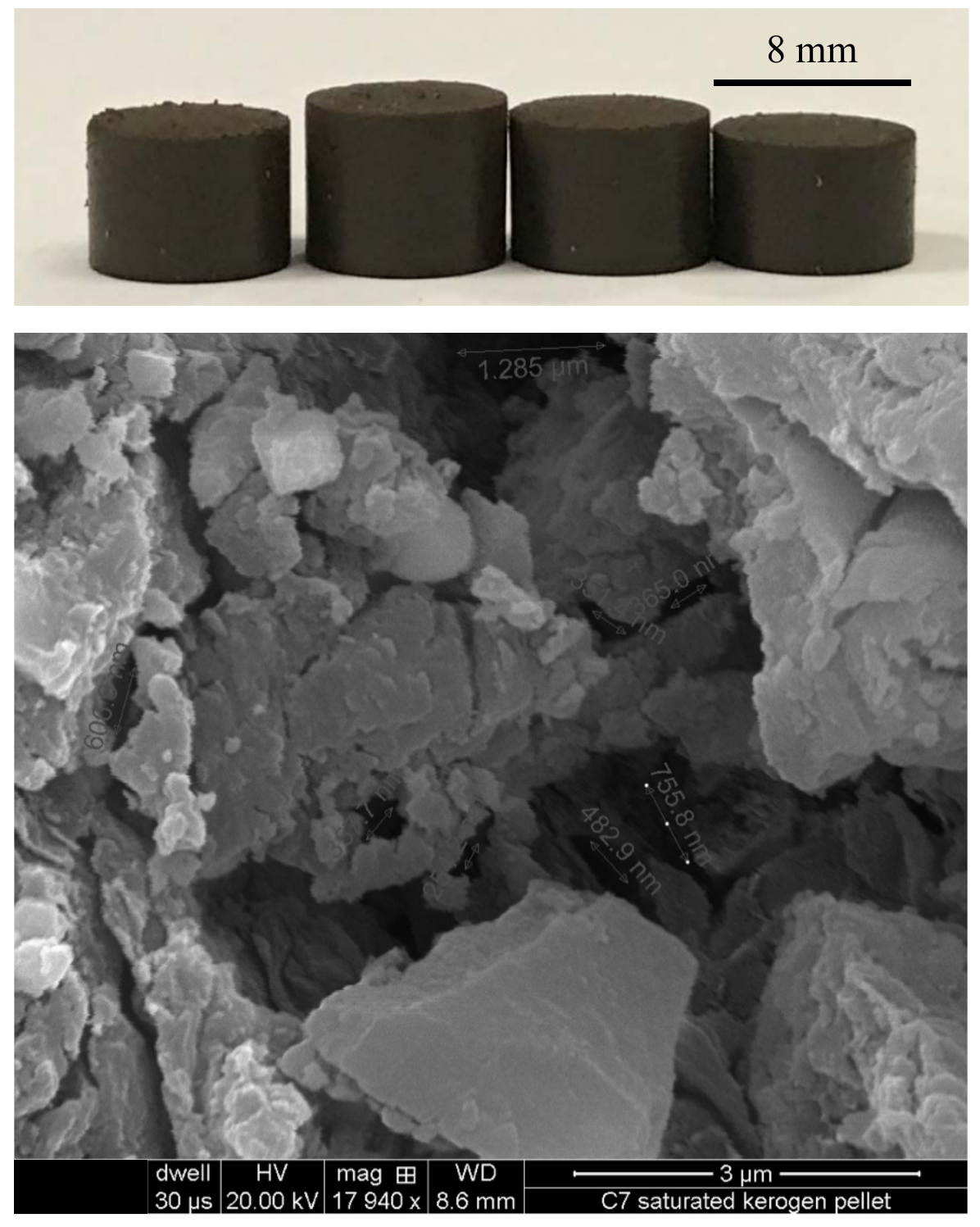} 
		\end{center}
		\vspace{-0.5cm}
		\caption{Picture of kerogen pellets (top) isolated from the Kimmeridge Shale. SEM (scanning electron microscope) image of the crushed pellets after heptane saturation (bottom). Images are taken from Ref. \citenum{singer:petro2016} and \citenum{singer:fuel2020}.} \label{fg:Pellets}
	\end{figure}

	The kerogen isolation procedure can be found in Ref.~\citenum{singer:fuel2020}, resulting in a high organic-matter content of $TOC \simeq 68$ wt\% (or $\simeq 84 $ vol\% equivalently). The dry kerogen powder was then pelletized in an 8 mm diameter pellet press with an (axial) pelletizing stress of 51,000 psi. An example of the pellets is shown in Fig. \ref{fg:Pellets}, where the total dry mass was $\simeq 1$ g. Also shown in Fig. \ref{fg:Pellets} is an SEM (scanning electron microscope) image of the kerogen grains and the inter-granular pore-space. The pellets were then vacuum-saturated by applying a high vacuum for 2 hours followed by saturation with $n$-heptane at 5000 psia hydrostatic pressure for 4 days. Saturation with heptane resulted in $\simeq$50\% swelling of the bulk volume of the pellets, where the swelling took place mostly along the axial direction, i.e., in the direction of the axial stress during pelletization.
	
	The NMR data were acquired on an Oxford Instruments GeoSpec2 rock-core analyzer at a resonance frequency of $f_0$ = 2.3 MHz (similar to wireline NMR logging tools) using a customized 18 mm probe. The pellets were placed in a sealed glass tube, and left for one hour in the spectrometer to equilibrate to 30$^{\circ}$C before data acquisition. The 2D $T_1$-$T_2$ data were acquired with 32 log-spaced inversion recovery steps ranging from 0.1 to 1000 ms for $T_1$ encoding, followed by a CPMG (Carr, Purcell, Meiboom, Gill) train with an echo spacing of $t_E$ = 0.1 ms for $T_2$ encoding. 
	
	The 2D maps were processed using Tiknonov regularization \cite{venkataramanan:ieee2002,song:jmr2002} with 120 log-spaced bins ranging from 10$^{-2}$ ms to 10$^{3}$ ms for both $T_1$ and $T_2$. The total porosity $\phi_T$ is reported in ``pu", i.e. porosity units, or percent of bulk volume, equivalently, where the NMR hydrogen index of $HI = $ 1.0028 for $n$-heptane is taken into account. 
	
	The cutoff separating the inter-granular heptane from the dissolved heptane is chosen to be $T_2 = 2$ ms (dashed vertical line). The inter-granular porosity $\phi_>$ above $T_2 > 2 $ ms, the intra-granular porosity $\phi_<$ below $T_2 < 2 $ ms, and the total NMR porosity $\phi_T = $ 44.5 pu are related by:
	\begin{align}
		&\phi_{T} = \phi_< + \phi_> \quad {\rm (pu)}\label{eq:phi_tot}\\
		&\phi_{\mu} = \frac{\phi_<}{1-\phi_>} \quad {\rm (vol\%)}, \label{eq:phi_micro}
	\end{align}
	and $\phi_{\mu} = $ 13.6 vol\% is the ``micro-porosity", i.e. the dissolved-heptane porosity of the kerogen grain in vol\%. The units of the $y$-axis of the 1D distributions $P(T_1)$ and $P(d)$ are in (vol\%/div), which stands for volume percent per ``division''. A division, defined as div $= \log_{10} (T_{1,i+1} )-\log_{10} (T_{1,i})$, is the logarithmic bin spacing, which is independent of index $i$. This unit convention ensures that a square distribution a decade wide and unit height has an area of 1 vol\%.
	
	Note that in addition to cross-relaxation effects in $T_1$ between all $^1$H's as a result of spin diffusion \cite{washburn:cmr2014,singer:jcp2018}, there is also diffusive coupling (for both $T_1$ and $T_2$) between inter-granular and dissolved heptane as a result of molecular diffusion \cite{singer:fuel2020}. These two phenomena, i.e., spin diffusion and molecular diffusion, have distinct physical origins, though they are often confused with each other.

\newpage

\providecommand{\latin}[1]{#1}
\makeatletter
\providecommand{\doi}
{\begingroup\let\do\@makeother\dospecials
	\catcode`\{=1 \catcode`\}=2 \doi@aux}
\providecommand{\doi@aux}[1]{\endgroup\texttt{#1}}
\makeatother
\providecommand*\mcitethebibliography{\thebibliography}
\csname @ifundefined\endcsname{endmcitethebibliography}
{\let\endmcitethebibliography\endthebibliography}{}